\documentclass{aa}  
\usepackage[varg]{txfonts}
\usepackage{graphicx}
\usepackage{epsfig}
\usepackage{color}
\usepackage{amssymb}
\usepackage{txfonts}
\usepackage{lscape}
\usepackage{float}
\usepackage{amsmath}
\usepackage{color}
\usepackage{longtable}
\usepackage{journal_names}
\usepackage{rotating}
\usepackage{pdflscape}
\usepackage{dcolumn}
\usepackage{caption}
\usepackage{subcaption}
\usepackage[nameinlink]{cleveref}

\usepackage{color,soul}
\usepackage{amssymb}
\usepackage{amsmath}
\usepackage{bm}
\usepackage{lscape}
\usepackage{float}
\usepackage{enumitem}
\usepackage{mathtools}
\DeclareMathOperator{\MyProd}{\scalebox{1.4}{$\mathrm{I\kern-0.2ex I}$}}

\usepackage{subcaption}
\usepackage{caption}
\usepackage{xcolor}
\usepackage[export]{adjustbox}
\DeclareCaptionLabelFormat{continued}{Figure \ref{postage}}
\newcommand{\HI}{\mbox{\normalsize H\thinspace\footnotesize I}}
\newcommand{\MSUN}{${\rm M}_\odot$}
\newcommand{\kms}{{\,km\,s$^{-1}$}} 
\newcommand{\dg}{^{\circ}} 
\def\arcmin{\hbox{$^\prime$}}
\def\arcsec{\hbox{$^{\prime\prime}$}}

\begin{document} 

\title{GASP XXVI. HI gas in jellyfish galaxies}

\subtitle{The case of JO201 and JO206.}
\author{M. Ramatsoku\inst{1,}\inst{2}\fnmsep\thanks{m.ramatsoku@ru.ac.za} \and  P. Serra \inst{2} \and B. M. Poggianti\inst{3} \and A. Moretti\inst{3} \and M. Gullieuszik\inst{3} \and D. Bettoni\inst{3} \and T. Deb\inst{4} \and A. Franchetto\inst{3,15} \and J. H. van Gorkom\inst{5} \and Y. Jaff\'{e}\inst{6} \and S. Tonnesen\inst{7} \and M. A. W Verheijen\inst{4} \and B. Vulcani\inst{3} \and L. A. L. Andati\inst{1} \and E. de Blok\inst{11,12,4} \and G. I. G. J\'{o}zsa\inst{8,1,9} \and P. Kamphuis\inst{14} \and D. Kleiner\inst{2} \and F. M. Maccagni\inst{2} \and S. Makhathini\inst{1} \and D. Cs. Moln\'{a}r\inst{2} \and A. J. T. Ramaila\inst{8} \and O. Smirnov\inst{1,8} \and K. Thorat\inst{13}}

\institute{Department of Physics and Electronics, Rhodes University, PO Box 94, Makhanda, 6140, South Africa 
         \and
             INAF- Osservatorio Astronomico di Cagliari, Via della Scienza 5, I-09047 Selargius (CA), Italy
         \and
            INAF- Osservatorio Astronomico di Padova, Vicolo dell'Osservatorio 5, I-35122 Padova, Italy
         \and
           Kapteyn Astronomical Institute, University of Groningen, Landleven 12, 9747 AV Groningen, The Netherlands
         \and
           Department of Astronomy, Columbia University, Mail Code 5246, 550 W 120th Street, New York, NY 10027, USA
         \and
           Instituto de F\'{i}sica y Astronom\'{i}a, Universidad de Valpara\'{i}so, Avda. Gran Bretana 1111 Valpara\'{i}so, Chile
         \and 
           Center for Computational Astrophysics, Flatiron Institute, 162 5th Ave, New York, NY 10010, USA
         \and 
           South African Radio Astronomy Observatory, 2 Fir Street, Black River Park, Observatory, 7925, South Africa
         \and
           Argelander-Institut fur Astronomie, Auf dem Hugel 71, D-53121 Bonn, Germany
         \and 
           National Centre for Radio Astrophysics, Tata Institute of Fundamental Research,  Postbag 3, Ganeshkhind, Pune 411 007, India
          \and 
           ASTRON, Netherlands Institute for Radio Astronomy, Oude Hoogeveensedijk 4, 7991 PD, Dwingeloo, The Netherlands
          \and 
           Department of Astronomy, University of Cape Town, Private Bag X3, Rondebosch 7701, South Africa
          \and
           Department of Physics, University of Pretoria, Hatfield, Pretoria, 0028, South Africa
          \and 
           Ruhr University Bochum, Faculty of Physics and Astronomy, Astronomical Institute, 44780 Bochum, Germany 
          \and 
          Dipartimento di Fisica e Astronomia ``Galileo Galilei'', Universit\`{a} di Padova, vicolo dell'Osservatorio 3, IT-35122, Padova, Italy.
           }

 \date{Received 19 February 2020; accepted 02 June 2020}

 
  \abstract{We present atomic hydrogen (\HI) observations with the Jansky Very Large Array of one of the jellyfish galaxies in the GAs Stripping Phenomena sample, JO201. This massive galaxy (M$_{\ast} =$ 3.5 $\times$ 10$^{10}$ \MSUN) is falling along the line-of-sight towards the centre of a rich cluster (M$_{200} \sim 1.6 \times 10^{15}$\MSUN, $\sigma_{cl} \sim\ 982 \pm 55$\kms) at a high velocity $\geq$3363\,\kms. Its H$\alpha$ emission shows a  $\sim$40\,kpc tail, which is closely confined to its stellar disc and a $\sim$100\,kpc tail extending further out. We find that \HI\ emission only coincides with the shorter clumpy H$\alpha$ tail, while no \HI\ emission is detected along the $\sim$100\,kpc H$\alpha$ tail. In total, we measured an \HI\ mass of M$_{\rm HI} = 1.65 \times 10^{9}$ \MSUN, which is about 60\% lower than expected based on its stellar mass and stellar surface density. We compared JO201 to another jellyfish in the GASP sample, JO206 (of a similar mass but living in a ten times less massive cluster), and we find that they are similarly \HI-deficient. Of the total \HI\ mass in JO201, about 30\% lies outside the galaxy disc in projection. This \HI\ fraction is probably a lower limit since the velocity distribution shows that most of the \HI\ is redshifted relative to the stellar disc and could be outside the disc. The global star formation rate (SFR) analysis of JO201 suggests an enhanced star formation for its observed \HI\ content. The observed SFR would be expected if JO201 had ten times its current \HI\ mass. The disc is the main contributor of the high star formation efficiency at a given \HI\ gas density for both galaxies, but their tails also show higher star formation efficiencies compared to the outer regions of field galaxies. Generally, we find that JO201 and JO206 are similar based on their \HI\ content, stellar mass, and star formation rate. This finding is unexpected considering their different environments. A toy model comparing the ram pressure of the intracluster medium (ICM) versus the restoring forces of these galaxies suggests that the ram pressure strength exerted on them could be comparable if we consider their 3D orbital velocities and radial distances relative to the clusters.}


\maketitle
%

\section{Introduction}
In the dense environment of galaxy clusters, the hydrodynamical interaction between galaxies' interstellar medium (ISM) and the intracluster medium (ICM) plays a key role in transforming galaxies from late types to early types. Among the several types of interactions discussed in the literature (e.g. \citealp{Gunn1972}, \citealp{Cowie1977}, \citealp{Nulsen1982}), the ram pressure exerted by the ICM on galaxies' ISM can be very efficient at removing gas from galaxies as well as affecting their star formation activity.
The degree of this effect may vary depending on the strength of the ram pressure exerted on a galaxy and its physical properties. 

In some cases, the diffuse atomic hydrogen (\HI) gas is only partially removed or displaced from the stellar disc (\citealp{Boselli1997}, \citealp{Vollmer2001}, \citealp{Chung2009}, \citealp{Scott2010}, \citealp{Steinhauser2016}). However, the dense molecular gas cloud may remain unperturbed after the diffuse gas has been stripped and continue to form stars at a low rate. In this case, the star formation (SF) essentially stops as soon as molecular gas is consumed and not replenished \citep{Abramson2014}.

In other cases, the ram pressure is so severe (e.g. \citealp{Jaffe2018}; \citealp{Quilis2000}) that galaxies are often observed to be \HI\ gas deficient and exhibit asymmetric morphologies (e.g. \citealp{Boselli2006}, \citealp{Fumagalli2014}, \citealp{McPartland2016}, \citealp{Gavazzi2018}). These galaxies sometimes show an increased star formation rate even to the point of becoming temporarily brighter than the bright central galaxies of their host clusters (e.g. \citealp{Owen2006}, \citealp{Cortese2007}, \citealp{Owers2012}, \citealp{Ebeling2014}, \citealp{George2018}, \citealp{Vulcani2018}). This implies that the compression of the ISM could possibly induce starbusts.

Moreover, the stripped gas may form knots of star formation in the wake of the galaxy. Examples of these types of cases have been observed in the so called `jellyfish' galaxies (e.g. \citealp{Yoshida2008}, \citealp{Hester2010},  \citealp{Yagi2010},  \citealp{Yagi2010}, \citealp{Fumagalli2014}, \citealp{Boselli2016}, \citealp{Fossati2016}, \citealp{Poggianti2019}, \citealp{Cramer2019}). The stripped tails of some of these galaxies are detected in H$\alpha$ and Ultraviolet (UV) (\citealp{Cortese2006}, \citealp{Sun2007}, \citealp{Kenney2008}, \citealp{RSmith2010}, \citealp{Ebeling2014}, \citealp{George2018}). 

Detailed observations of these galaxies, which include their gaseous tails, are important to gain a full understanding of the ram pressure effect on the star formation efficiency of jellyfish galaxies. While some studies have shown that these galaxies show an overall enhanced SF (e.g. \citealp{Boissier2012}, \citealp{Vulcani2018} \citealp{Ramatsoku2019}), others find a reduced star formation due to ram pressure stripping (\citealp{Vollmer2012}, \citealp{Jachym2014}, \citealp{Verdugo2015}). 
 
These reports make it imperative to gather large multi-wavelength samples and to conduct detailed studies of these objects (jellyfish galaxies) that are undergoing a transformation in the galaxy cluster environments. 
The GAs Stripping Phenomena survey (GASP; \citealp{Poggianti2017}) has been performed in an effort to conduct this type of a study. Through GASP, a statistically significant sample of jellyfish galaxies has been compiled in nearby clusters ($z = 0.04 - 0.07$) from the WIde-field Nearby Galaxy-cluster Survey (WINGS; \citealp{Fasano2006}) and OmegaWINGS (\citealp{Cava2009}, \citealp{Varela2009}, \citealp{Gullieuszik2015}, \citealp{Moretti2017}). The current GASP operational definition of a jellyfish galaxy is one that exhibits a one-sided tail of debris material whose H$\alpha$ emission has a length comparable or greater to the diameter of the stellar disc. The optically selected jellyfish candidates were observed with the Multi Unit Spectroscopic Explorer (MUSE) Integral Field spectrograph on the Very Large Telescope (VLT) \citep{Poggianti2016}. This sample covers a wide range of jellyfish morphological asymmetries and masses in different environments. Within the GASP context, the aim is to understand the evolution of the distribution and efficiency of star formation in ram pressure stripped galaxies by studying all phases of the gas in relation to the stellar population properties of these galaxies.

The large radial distribution of the \HI\ gas in galaxies and its sensitivity to mild ram pressures makes it an excellent tracer of gas removal processes in cluster environments (\citealp{Haynes1984}, \citealp{Bravo-Alfaro1997}, \citealp{Gavazzi2008}, \citealp{Kapferer2009}, \citealp{Chung2009}, \citealp{Abramson2011}, \citealp{Jaffe2015}, \citealp{Yoon2017}). Therefore to fully understand the physics and the effect of ram pressure stripping on the GASP galaxy sample it is necessary to examine the \HI\ content and distribution.  This is what the work presented in this paper focuses on.

We have conducted a follow up on a few GASP galaxies with the Jansky Very Large Array (JVLA; \citealp{Perley2009}) to study their \HI\ properties. \citet{Deb2020} investigated whether ram pressure stripping might be triggering the Active Galactic Nucleus (AGN) in JO204 by funnelling \HI\ towards the centre of the galaxy. In \citet{Ramatsoku2019} we examined the \HI\ gas stripping and star formation activity of JO206, which resides in the poor galaxy cluster, IIZw108 (M$_{200} \sim 2 \times 10^{14}$\MSUN\, $\sigma_{cl} \sim\ 575 \pm 33$\kms; \citealp{Biviano2017}). We found that the star formation efficiency was higher in JO206 than in galaxies with similar stellar and \HI\ mass.

In this paper we present a study of the neutral gas phase of the jellyfish galaxy, JO201 ($\alpha_{J2000}, \delta_{J2000}$ = 00:41:30.30, -09:15:45.98; \citealp{Gullieuszik2015}). This galaxy resides within the rich galaxy cluster, Abell\,85 located at a redshift of $z = 0.05586$ \citep{Moretti2017}. 
JO201 is undergoing ram pressure stripping along the line of sight in this cluster and exhibits a one-sided tail of stripped material seen at the optical wavelength (\citealp{Poggianti2016}, \citealp{Bellhouse2017}, \citeyear{Bellhouse2019}). For this study we aim to compare and contrast the \HI\ properties and star formation activities of these two galaxies in consideration of their different environments.

The paper is organised as follows; in section\,\ref{PropJ201} we give a summary of the properties of JO201 as well as an overview of the available multiwavelength observations. We outline how the \HI\ observations and data reduction were conducted for JO201 in Section\,\ref{sec:vlaObs}. In Section\,\ref{j201HIresults} we discuss the \HI\ distribution compared to the H$\alpha$ of this galaxy. We compare the \HI-deficiencies, global and resolved star formation rates as well as the environments of JO201 with those of JO206 in Section.\,\ref{comparisons}. All the analyses and discussions are summarised in Section\,\ref{summary}.

Throughout this paper we adopt a \citet{Chabrier2003} initial mass function (IMF) and assume a $\Lambda$ cold dark matter cosmology with $\Omega_{\rm M} = 0.3, \Lambda_{\Omega} = 0.7$ and a Hubble constant, H$_{0}$ = 70 \kms\ Mpc$^{-1}$ .

\section{Properties of the JO201 galaxy}\label{PropJ201}
JO201, is a spiral galaxy with an AGN at its centre (\citealp{Poggianti2017nat}, \citealp{Radovich2019}) and a stellar mass of $3.6 \times10^{10}$ \MSUN\ \citep{Vulcani2018}. It is located within the massive and dynamically complex galaxy cluster, Abell\,85 (A\,85) in the WINGS/OmegaWINGS sample (\citealp{Fasano2006}, \citealp{Cava2009}, \citealp{Varela2009}, \citealp{Gullieuszik2015}, \citealp{Moretti2017}). A\,85 has a total mass of M$_{200} \sim 1.6 \times 10^{15}$\MSUN\  and a velocity dispersion of $\sigma_{cl} \sim\ 982 \pm 55$\kms\ \citep{Moretti2017}. Within this cluster, JO201 lies at a projected radial distance of 360\,kpc from the central brightest cluster galaxy and has a particularly high line-of-sight velocity of about $3.4\sigma_{cl}$ relative to the cluster's systemic velocity \citep{Bellhouse2017}.

JO201 is experiencing ram pressure stripping due to the ICM of A\,85 \citep{Bellhouse2017}. Its effect is manifested in the form of two long one-sided tails visible in H$\alpha$ within the GASP survey and debris material as seen at optical wavelengths in the WINGS/OmegaWINGS surveys (\citealp{Varela2009}, \citealp{Gullieuszik2015}, \citealp{Moretti2017}). \citet{Bellhouse2017} performed a quantitative analysis of the strength of ram pressure exerted on JO201 by the ICM of A\,85 using the \citet{Gunn1972} description (P$_{\rm ram}$ = $\rho_{\rm \rm \scriptscriptstyle ICM}v^{2}$), and calculated a ram pressure strength of P$_{\rm ram} =$ $2.7 \times 10^{-11}$ Nm$^{-2}$ at the projected distance of the galaxy from the cluster centre. The exerted ram pressure was then compared to its anchoring force, suggesting that more than 40\% of the total gas (estimated from H$\alpha$) had been removed through ram pressure stripping (see \citealp{Bellhouse2017}, \citeyear{Bellhouse2019} for the full analysis).

\begin{figure}
\hspace*{-0.5cm} 
 \includegraphics[width=90mm, height=100mm]{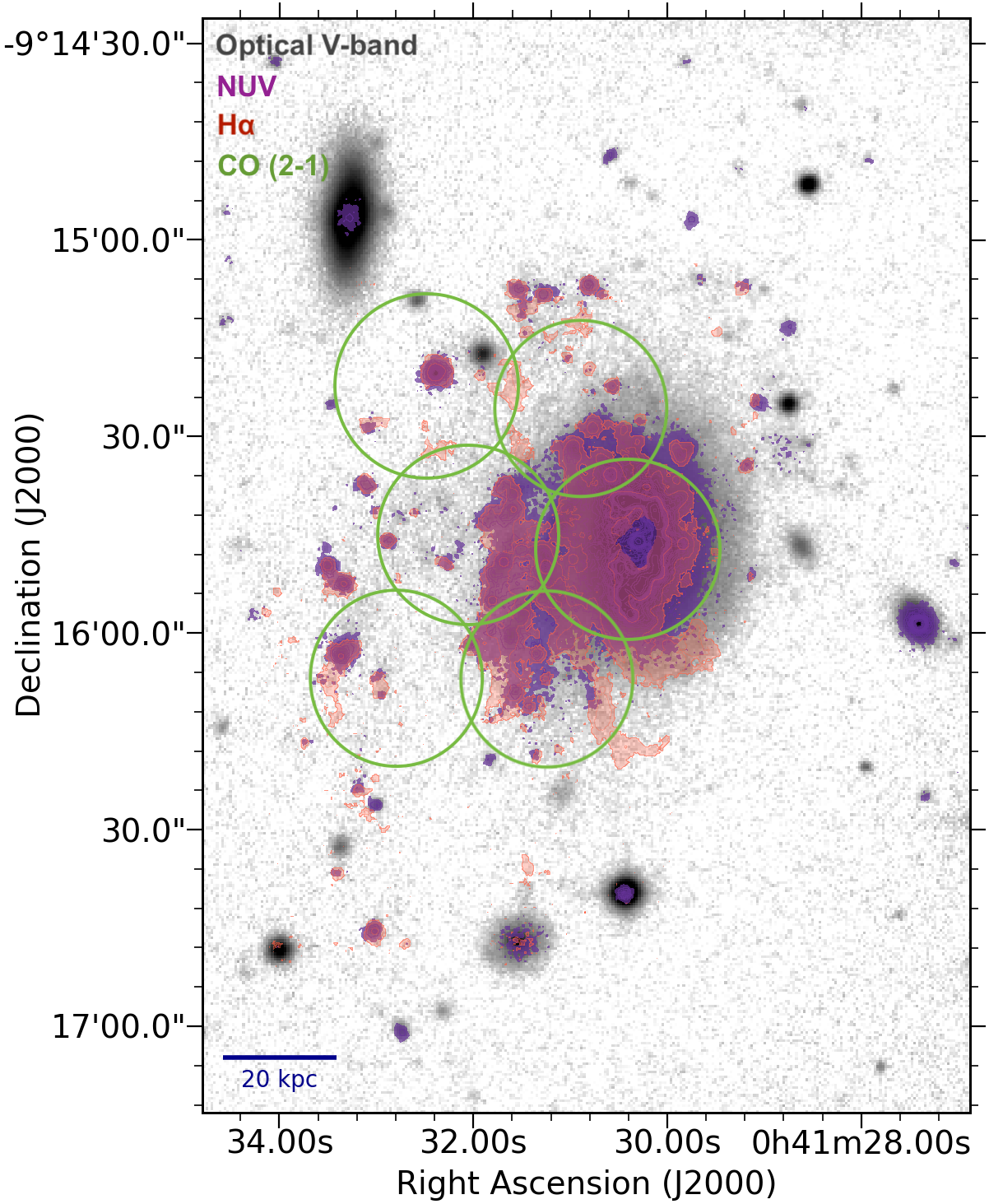}
    \caption{A multiwavelength composite image illustrating all the data available for JO201. The NUV data is shown in violet overlaid on an optical V-band image in grayscale from WINGS. In red we overlay the H$\alpha$ emission from MUSE. The $^{12}$CO(2-1) observations are illustrated by the approximate FWHM of the APEX beam.}\label{multiwavelengthFig}
 \end{figure}

Although this galaxy has lost about half of its gas due to ram pressure stripping, it is still forming new stars as inferred from its H$\alpha$ MUSE data, UV imaging from GALEX and ASTROSAT, and CO data from APEX (\citealp{Bellhouse2019}, \citealp{George2018}, \citealp{Venkatapathy2017}, \citealp{Moretti2018}). These data are illustrated in Fig.\,\ref{multiwavelengthFig}. The total SFR estimated from the H$\alpha$ emission after masking the AGN and correcting for stellar absorption and dust extinction\footnote{The dust correction was estimated from the Balmer decrements measured from the MUSE spectra.} as in Poggianti et al. (2017b), is 6.1 \MSUN/yr \citep{Vulcani2018}.

An analysis of the star formation history shows that the stellar disc has stars of all ages, while the stripped tail only comprises young stars formed $< 6 \times 10^{8}$ years ago (see Fig.\,9 by \citealp{Bellhouse2019}). These young stars have likely formed in-situ out of the ram pressure stripped ISM, thus strongly implying the presence of cold gas and, therefore, \HI\ both in the tail and disc. This measurement motivated the \HI\ observations described in Sec.\,\ref{sec:vlaObs}.

\section{HI observations and data processing}\label{sec:vlaObs} %
The \HI\ data were obtained from observations with the JVLA, project code VLA/17A-293 (PI: B. Poggianti). JO201 was observed in July 2017 for a total of 16 hours on source and 4 hours on calibrators. The observations were conducted in the C-configuration resulting in an angular resolution of $\sim$15\arcsec. They covered a total bandwidth of 32 MHz, centred at 1360.2103 MHz, which was divided into 1024 channels of width 31.25 kHz each (6.56\,\kms\ for \HI\ at $z$ = 0). 

The uv-data were processed using the \textsc{caracal} pipeline\footnote{https://caracal.readthedocs.io} which is currently being developed. Using this pipeline we followed the same data reduction steps outlined in \citet{Ramatsoku2019}. As a brief summary, an \HI\ data cube with a field of view of 0.7 deg$^2$ was produced using a pixel scale of 5\arcsec\ with natural weighting (\textit{Briggs} robustness parameter $=$ 2). We chose this weighting scheme to optimise surface-brightness sensitivity. The resulting \HI\ cube has a Gaussian restoring Point Spread Function(PSF) Full Width at Half Maximum (FWHM) of 16\arcsec\ $\times$ 25\arcsec\ and a position angle (PA) of 174$\dg$. The rms noise level is $\sigma \approx$ 0.37 mJy beam$^{-1}$ per 6.56\,\kms-wide channel. The observational setup and noise level results in an \HI\ column density sensitivity of $4 \times 10^{19}$ atoms cm$^{-2}$ ($3\sigma$ over a line width of 30 \kms). For a detailed description of the \HI\ data reduction see \citet{Ramatsoku2019}.

\section{The distribution of H\textsc{i} in JO201 and comparison with H$\alpha$}\label{j201HIresults}
The \HI-line emission of JO201 was extracted from the image cube using SoFiA \citep{Serra2015}. Within the SoFiA \HI\ extraction mask we measure a total \HI\ mass of M$_{\rm HI} = 1.65 \times 10^{9}$ \MSUN, assuming the galaxy is at the redshift of the cluster, $z = 0.05586$, D $=$  239\,Mpc \citep{Moretti2017}. 

\begin{figure*}[h!]
\hspace*{-0.5cm} 
 \includegraphics[width=188mm, height=85mm]{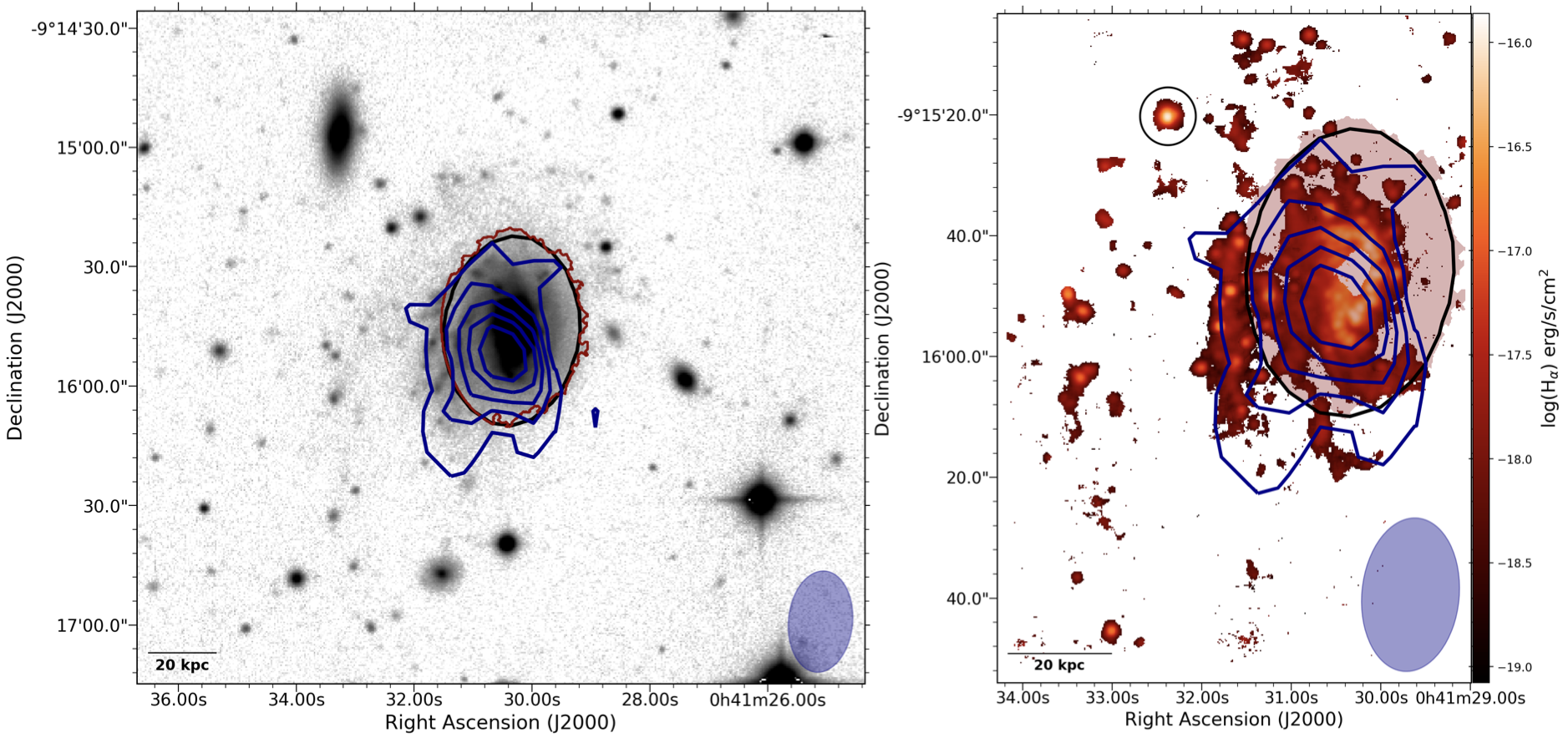}
    \caption{In the left panel the VLA H\textsc{i} column density contours are overlaid on the V-band image of JO201 from WINGS. Contour levels are drawn at column densities of 4, 8, 12, ... $\times$ 10$^{19}$ atoms/cm$^{2}$. The FWHM beam size of $16\arcsec\ \times 25\arcsec$ is indicated by the blue ellipse. The red contours represent the stellar disc and the black contour is the H\textsc{i} disc (see text for a definition). The right panel is the H$\alpha$ emission with H\textsc{i} column density contours (same levels as the left panel) overlaid. The red opaque patch is the stellar disc and the black contour is the H\textsc{i} disc. The black circle is the region over which we estimate the H\textsc{i} flux density as described in the text.}\label{HIoverVband}
 \end{figure*}

The left panel of Fig.\,\ref{HIoverVband} shows the spatial distribution of the \HI\ emission in JO201. The blue contours are from the \HI\ intensity map (i.e. moment 0). They are overlaid on an optical V-band image from WINGS/OmegaWINGS. The red contour outlines the stellar disc defined from the MUSE image using the continuum at the H$\alpha$ wavelength, with the isophote fit at surface brightness of 1$\sigma$ above the average sky background level (Gullieuszik et al., submitted).

The black contour represents the expected \HI\ disc size if all the detected \HI\ were distributed on a regular disc and followed standard \HI\ scaling relations. We calculated its size at the $4 \times 10^{19}$ atoms cm$^{-2}$ \HI\ column density sensitivity of our JVLA data taking into account the inclination angle derived from the stars following the methods described in \citet{Ramatsoku2019}. To make this calculation we exploited the tight correlation between \HI\ diameters and masses which is parametrised as $\mathrm{log (D_{HI}/kpc) = 0.51\,log(M_{HI}/M_{\odot}) -  3.32}$, where the \HI\ diameter, D$_{\rm HI}$ is defined at the \HI\ surface density of 1 \MSUN\ pc$^{-2}$ (column density $1.25 \times 10^{20}$ atoms cm$^{-2}$; \citealp{Wang2016}). From this relation we compute an \HI\ diameter of D$_{\rm HI}$  = 23\,kpc for the unperturbed JO201 model. We then used the \citet{Martinsson2016} \HI\ profile formulation, $\mathrm{\Sigma_{\rm H\textsc{i}}(R) = \Sigma_{\rm H\textsc{i}}^{\rm max}.e^{\frac{-(R - R_{\Sigma,\rm max})^{2}}{2\sigma_{\Sigma}^{2}}}}$ to determine the radial surface density. In this formula R$_{\Sigma,\rm max}$ and $\sigma_{\Sigma}$ are fixed to 0.2D$_{\rm HI}$ and 0.18D$_{\rm HI}$, respectively. The free parameter is $\Sigma_{\rm H\textsc{i}}^{\rm max}$, which we set to 0.4 \MSUN pc$^{-2}$ such that $\Sigma_{\rm HI}$(D$_{\rm HI}$/2) = 1 \MSUN pc$^{-2}$. With the above correlations and assuming the \HI\ observational conditions of JO201 (see Sec.\,\ref{sec:vlaObs}) we used the \textsc{3d-barolo} package \citep{DiTeodoro2015} to project the unperturbed \HI\ distribution of JO201 onto the sky with the correct inclination and PA. 

The model \HI\ disc has a radius of 24.6\,kpc at the $4 \times 10^{19}$ atoms cm$^{-2}$ \HI\ column density sensitivity of our data as shown in the left panel of Fig.\,\ref{HIoverVband}. We note that for JO201, the stellar and \HI\ disc are similar, therefore throughout the paper we do not differentiate between the two, unless stated otherwise.

We used the projected radius derived from the model (i.e. 24.6 kpc) as a divider between inner (disc) and outer galaxy regions, and find that of the total \HI\ mass measured $1.22 \times 10^{9}$ \MSUN\ is inside the galaxy disc with $\sim0.43 \times 10^{9}$ \MSUN\ (26\%) lying in the outer regions in projection. Below we revisit this analysis based on the velocity distribution as well.

The projected spatial distribution of the \HI\ and H$\alpha$ show moderately different distributions (e.g. Fig.\,\ref{HIoverVband}, right panel). \HI\ overlaps almost entirely with the H$\alpha$ emission within the galaxy disc. It also has a minor offset of $\sim5$\,kpc from the optical centroid in the direction of the stripped tail and shows a compression on the west side of the emission which is also seen in H$\alpha$. All these features are hallmarks of ram pressure stripping which we are seeing manifest in the \HI\ distribution. Both the one-sided compression to the west and the bulk offset towards east suggest that the motion of JO201 within A\,85 might have a non-negligible component on the plane of the sky.
Outside the galaxy disc, the overlap between \HI\ and H$\alpha$ tail is only seen along the shorter clumpy H$\alpha$ tail that is closely confined to the disc on the east side. 

The non-detection of \HI\ along the outer H$\alpha$ tail might be due to the small size of the gas clumps (at least judging from the H$\alpha$ image) compared to the \HI\ PSF. As an example, let us consider the brightest H$\alpha$ blob of the outer tail (black circle in the right panel of Fig.\,\ref{HIoverVband}) and test whether it would be detectable in our \HI\ cube. We generously assume that the \HI\ surface density inside this 5-kpc-diameter blob is 5\,\MSUN/pc$^{2}$, which is a typical value of the disc of spiral galaxies but is, in fact, quite high for extraplanar \HI\ (see top-left panel of Fig.\,8 in \citealp{Bigiel2008}). We further assume that the entire \HI\ signal is spread over a velocity interval of 50\,\kms\ (twice the typical FWHM of the \HI\ line). With these assumptions, and given the 239 Mpc distance of JO201, the \HI\ flux density of the blob would be 0.14 mJy beam$^{-1}$. This is equal to a 1$\sigma$ level when smoothing our \HI\ cube to a channel width of 50\,\kms\ in order to match the width of the assumed \HI\ signal. Despite our generous assumptions, the blob would thus not be detectable in our data. Therefore, while we can rule out the presence of an extended (i.e. resolved by our \HI\ PSF) distribution of \HI\ in the outer H$\alpha$ tail down to $4 \times 10^{19}$ atoms cm$^{-2}$, we cannot rule out that some of the H$\alpha$ clumps much smaller than our \HI\ PSF contain \HI\ with relatively high column density.

\begin{figure}
   \centering
 \includegraphics[width=85mm, height=110mm]{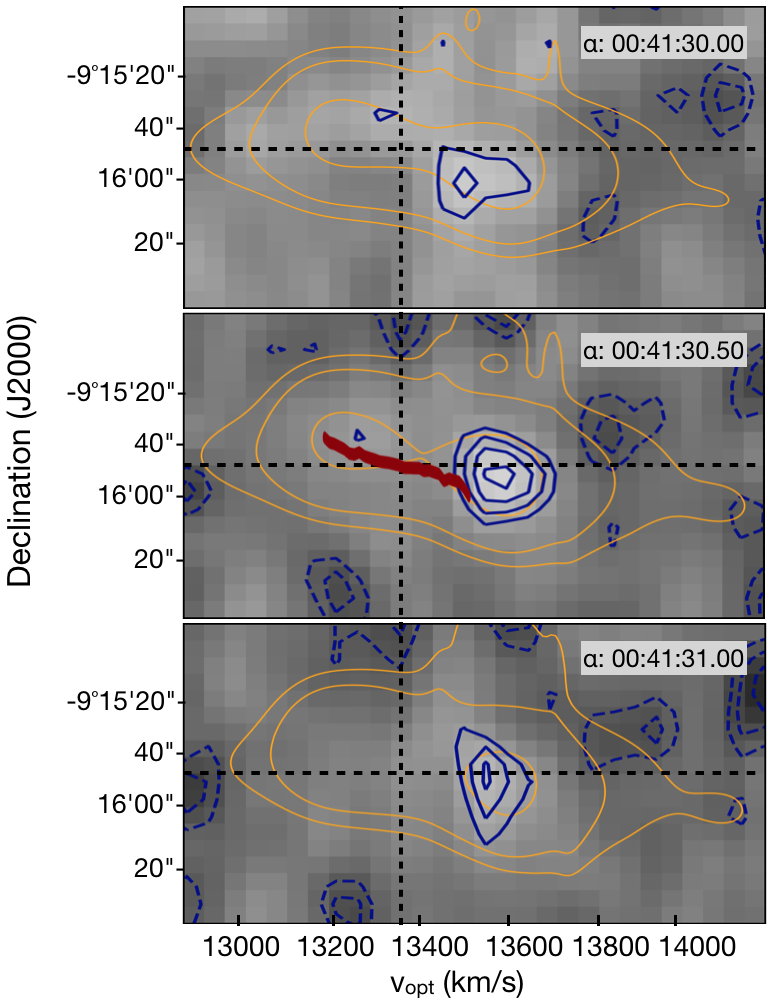}
    \caption{The position velocity diagrams (PVDs) extracted from the H\textsc{i} cube along the three 1.25\arcmin\ wide right ascension slices indicated in the top right corner. The H$\alpha$ emission convolved with the H\textsc{i} PSF is shown in orange. For reference, the red contour in the middle panel represents the stellar velocities extracted from \citet{Bellhouse2017} (no convolution to the HI angular resolution was performed). The H$\alpha$ contours are drawn at surface brightness levels of 1 $\times$ 10$^{n}$ erg/s/cm$^{2}$ where $n = -10,-9,-8 ...$ . The blue contours are the H\textsc{i} emission from the cube drawn at flux densities of 0.3, 0.4, 0.5 ... mJy beam$^{-1}$ and the negative H\textsc{i} contours are drawn at -0.3, -0.2 ... mJy beam$^{-1}$. The vertical dashed line is the systemic velocity of the galaxy and the horizontal line is the optical centre. Velocities in the PVDs are in the optical definition using the barycentric standard-of-rest.}\label{PVDs}
 \end{figure}

From the projected spatial distribution image it appears as though most of the detected \HI\ lies within the galaxy disc. However, in previous studies \citet{Bellhouse2017} analysed the kinematics of JO201 using MUSE data. In this work they reported that the galaxy had a smooth stellar disc with a co-rotating ionised H$\alpha$ gas within the inner 6\,kpc. However this stars-gas co-rotation is not seen in the outer stellar disc $>6$\,kpc. In these outer regions they found that the H$\alpha$ gas trails behind the undisturbed stellar disc due to the line-of-sight ram pressure stripping nature of this galaxy. In the tail (outside of the disc) the stellar kinematics can only be measured very close to the disc (and is decoupled from the gas, that is trailing behind), while further out in the tail there is no stellar disc to compare with. 
Here we use the position-velocity diagram (PVD) in Fig.\,\ref{PVDs} to examine the \HI\ gas kinematics and its relation to the stellar disc and H$\alpha$ emission. From this plot we find that as already reported (\citealp{Bellhouse2017}), H$\alpha$ is redshifted with respect to the systemic velocity. This displacement becomes more severe when we examine the \HI\ velocity distribution. In fact the \HI\ emission does not cross the systemic velocity at all. It appears that in addition to the already known east-west asymmetry (most of the \HI\ is to the east of the galaxy), there is also a clear velocity asymmetry. From these PVDs we deduce that it is possible that most of the \HI\ we detect is already outside of the disc -- contrary to what we see from the projected spatial distribution (e.g. Fig.\,\ref{HIoverVband}). The star-\HI\ velocity offset is $\gtrsim100$\,\kms, taking into account the marginal inclination correction of $\mathrm{1/sin(54\dg)}$. At this velocity offset it is possible for the \HI\ gas to be lifted above the stellar disc, which typically has scale heights of a few 100\,pc, because the gas can travel a distance of $\sim$1\,kpc in a short time-scale of $\sim$10\,Myr. This velocity offset therefore implies that much of the \HI\ we see might already be stripped and is part of the tail.

\begin{figure}
   \centering
 \includegraphics[width=90mm, height=92mm]{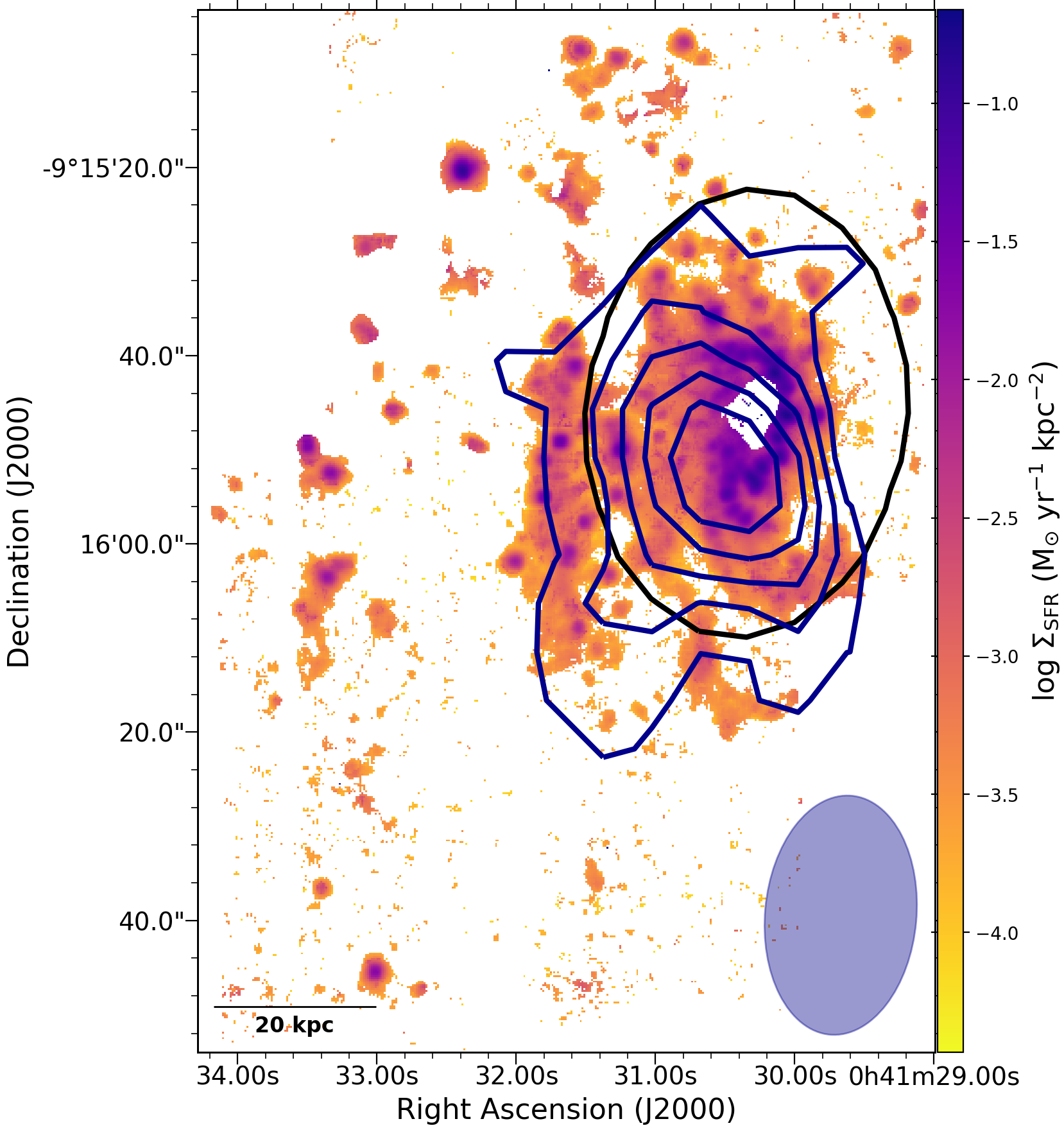}
    \caption{The star formation rate density map \citep{Bellhouse2019} with H\textsc{i} emission contours overlaid. The contours are at H\textsc{i} column density levels of 4, 8, 12, ... $\times$ 10$^{19}$ atoms/cm$^{2}$. The FWHM beam size of $25\arcsec \times 16\arcsec$ is shown by the blue ellipse.}\label{SFRDmap}
 \end{figure}

In addition to the H$\alpha$ and \HI\ distribution, we also examined \HI\ in relation to the star formation rate surface density ($\Sigma_\mathrm{SFR}$) of JO201 as shown in Figure\,\ref{SFRDmap}. We find a good agreement between peak $\Sigma_\mathrm{SFR}$ and \HI\ emission predominately in the projected disc of JO201. There is also a coincidence between the short \HI\ tails outside the galaxy disc and knots of high $\Sigma_\mathrm{SFR}$ in those regions.

\section{HI deficiency and SFR enhancement: JO201 and JO206}\label{comparisons}
In this section we are going to compare the \HI\ gas and star formation properties of JO201 with those of the other jellyfish galaxy with \HI\ data, JO206 in the GASP sample. 
JO206 is a massive galaxy with a stellar mass, M$_{\ast} =$ 8.5 $\times$ 10$^{10}$ \MSUN. It lies at a redshift of $z =$ 0.0513, near the centre of the low-mass (M$_{200} \sim 2 \times 10^{14}$ \MSUN) galaxy cluster IIZw108 with a velocity dispersion of $\sigma \sim575$\,\kms. This galaxy is characterised by a long tail ($\geq$90\,kpc) of ionised gas stripped away by ram pressure. 
In \citet{Ramatsoku2019} we analysed the \HI\ gas phase of JO206 and found a similarly long \HI\ tail extending in the same direction as the H$\alpha$ tail. Its total \HI\ mass is $3.2 \times 10^{9}$ \MSUN, of which $1.8 \times 10^{9}$ \MSUN\ (about 60\%) is in the stripped tail (see Fig.\, 3 in \citealp{Ramatsoku2019}).

\subsection{JO201 and JO206 HI deficiencies}
From the morphology of the \HI\ spatial distribution of JO201, it is evident that ram pressure stripping has affected the \HI\ content of this galaxy. 
A large fraction of the currently detected \HI\ is outside the disc, either in projection on the sky or in velocity (Sec.\,\ref{j201HIresults}). It is thus possible that JO201 used to have even more \HI, which is no longer detectable. Here we compute whether the total detected \HI\ mass of JO201 is significantly less than expected based on known \HI\ scaling relations and compare it with that of JO206. We do so using the scaling relations by \citet{Brown2015} obtained from spectral stacking of ALFALFA data. 

\begin{figure}
   \centering
 \includegraphics[width=85mm, height=90mm]{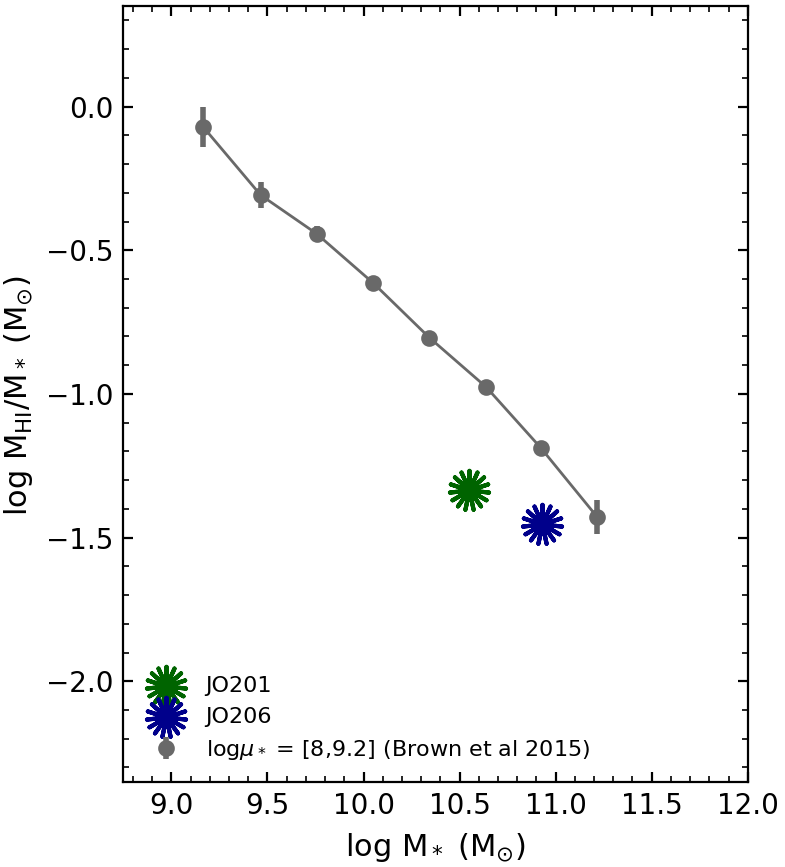}
    \caption{The average stacked H\textsc{i} fraction as function of the stellar mass \citep{Brown2015}. The relation is separated into galaxies with stellar surface brightness comparable to that of JO201 and JO206 within a factor of 4. The scatter in the mass bins is illustrate by the error bars. JO201 is represented by the green asterisk and JO206 by the blue asterisk.}\label{HIstack}
 \end{figure}

Fig.~\ref{HIstack} shows the scaling between \HI\ fraction M$_{\rm HI}/$M$_{\ast}$ and stellar mass $M_{\ast}$ for galaxies in the \citet{Brown2015} sample with a stellar surface mass density $\mu_\ast$ within a factor of 4 of the values measured for JO201 and JO206. This factor is comparable to the uncertainty in the $\mu_{\ast}$ values. JO201 with its  $\mu_{\ast} = 1.8 \times 10^{8}$ \MSUN\ kpc$^{-2}$ (log\,$\mu_{\ast}$ = 8.3) stellar surface density is represented by the green asterisks while JO206 with $\mu_{\ast} = 4.2 \times 10^{8}$ \MSUN\ kpc$^{-2}$ (log\,$\mu_{\ast}$ = 8.6) is denoted in blue.

Compared to the control sample, JO201 is approximately 0.4\,dex below the average \HI\ fraction. From the scaling relation the expected \HI\ mass for this galaxy is about $4.13 \times 10^{9}$\MSUN\ which indicates that JO201 is missing $\sim$60\% of its \HI\ mass and is therefore \HI\ deficient. This gas mass loss is comparable to that measured based on instantaneous ram pressure description in \citet{Bellhouse2017}. It is also similar to the \HI\ gas mass loss of JO206 which is 0.3\,dex below the average \HI\ fraction resulting in $\sim$50\% \HI\ deficiency.

We also determined the \HI\ deficiency of these galaxies using the parameter, Def$_{\rm HI} = \left<{\rm log}{\rm M}_{\rm HI}(\rm D,T)\right> - {\rm log}{\rm M}_{\rm HI} (\rm D,T)_{\rm obs}$ \citep{Haynes1984}, which defines the \HI-deficiency as the logarithmic difference between the observed \HI\ content and the expected value in isolated galaxies of the same linear size and morphology. The respective morphologies of JO201 and JO206 are Sa-Sab and Sb-Sbc \citep{Fasano2012} and their R$_{25}$ radii in the B-band are 17.9\arcsec\ and 24.95\arcsec. 
With this information we calculate Def$_{\rm HI}$= 0.44 for JO201 and 0.41 for JO206. The deficiencies of this order (about a factor of 2) are comparable with those we reported from the \HI\ scaling relation. However, we note that they are somewhat within the scatter of typical \HI\ scaling relations. Thus it is possible (although not necessary) that most of the \HI\ originally present in these galaxies is still detectable within or close to the galaxy.

\subsection{Global star formation activity}\label{sfrHIgals}
\begin{figure}
   \centering
 \includegraphics[width=90mm, height=70mm]{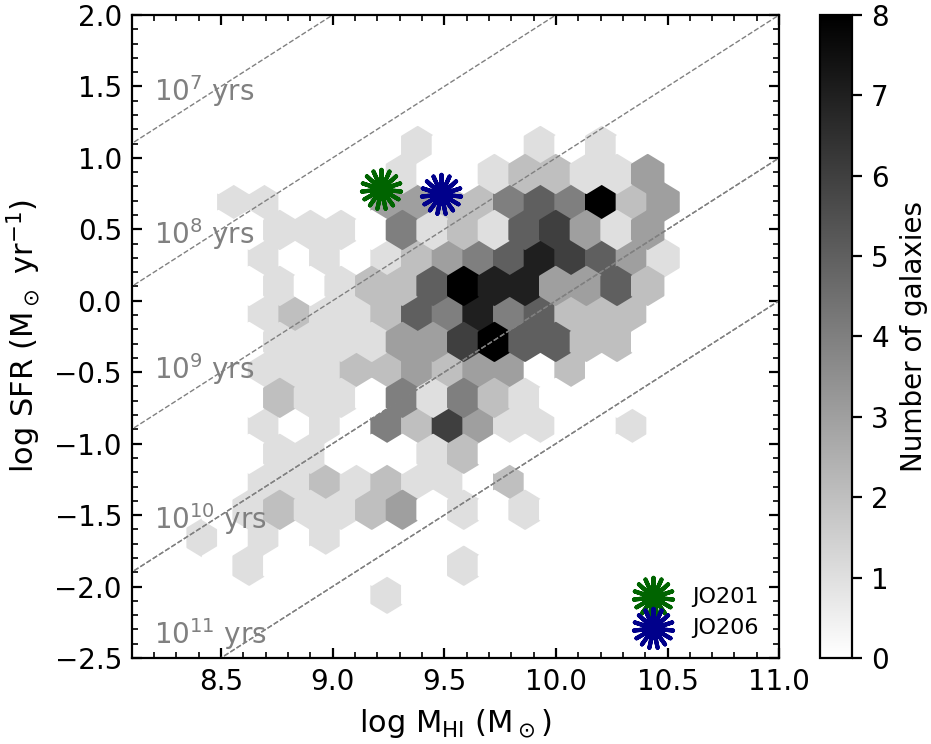}
    \caption{The H\textsc{i}-SFR scaling relation showing the star formation rate vs H\textsc{i} mass for galaxies with the same stellar masses as JO201 and JO206. Grey hexagons are galaxies from the control sample (GASS) and the green and blue asterisk represent the JO201 and JO206 galaxies, respectively. The dotted lines indicate the fixed H\textsc{i} depletion times in yr.}\label{SFRHI}
 \end{figure}

We investigate how the gas loss discussed in the previous section may have affected the global star formation activity of JO201, and how this compares to JO206. We begin by preparing a control sample of field galaxies with the same stellar mass as JO201 and JO206 obtained from the GALEX Arecibo SDSS Survey (GASS; \citealp{Catinella2010}). We also collected a sample of galaxies from The \HI\ Nearby Galaxy Survey (THINGS; \citealp{Bigiel2008}, \citealp{Bigiel2010}) which we use in section\,\ref{spatiallyres} to investigate the spatially resolved \HI-SFR relation.
In Fig.\,\ref{SFRHI} we show the global \HI-SFR scaling relation for the GASS sample (\citealp{Saintonge2016}; see also \citealp{Doyle2006} and \citealp{Huang2012})

JO201 (SFR = 6.1 \MSUN yr$^{-1}$) is located approximately 0.6\,dex above the average M(H\textsc{i)}-SFR relation similarly to JO206 (SFR = 5.6 \MSUN yr$^{-1}$). Both galaxies lie at the edge of the distribution of the control sample, and appear to have a higher star formation rate than galaxies of similar stellar mass given their observed \HI\ mass. We note however, that as shown in Fig.\,\ref{SFRHI} there is a large uncertainty associated with the exact SFR enhancement due the large scatter of this \HI-SFR scaling relation (see \citealp{Saintonge2016}). Nevertheless similarly to JO206, we argue that the fact that JO201 is located close to the edge of this distribution indicates that an enhancement of this galaxy's SFR or of the \HI\ conversion to H$_{\rm 2}$ might have taken place. The total SFR measured for JO201 would be expected if a galaxy with the same stellar mass had an \HI\ mass an order of magnitude above the measured M$_{\rm HI}$ ($\sim$ $2 \times 10^{10}$ \MSUN).
Both galaxies have short \HI\ depletion time scales through star formation $\tau_{d}$(HI)  = M$_{\rm HI}$/SFR.  For JO201 $\tau_{d}$(HI) = 0.27\,Gyr and for JO206 $\tau_{d}$(HI) = 0.54\,Gyr. These \HI\ depletion timescales are much shorter than the typical $\sim$2\,Gyr of normal disc galaxies \citep{Leroy2008}. \\ 

\subsection{Resolved star formation activity}\label{spatiallyres}
Having conducted a global analysis of the star formation activity using the \HI-SFR scaling relations, we found that compared to other `normal' disc galaxies in the field, JO201 and JO206 have enhanced star formation rates for their observed \HI\ content. In this section we study this enhancement in a spatially resolved way to locate exactly where in the galaxy stars are forming more efficiently. This method has been used effectively to study the star formation efficiency in galaxy discs and tails (\citealp{Bigiel2008}, \citealp{Bigiel2010}, \citealp{Boissier2012}, \citealp{Ramatsoku2019}).

We use the same procedure described in \citet{Ramatsoku2019} which we briefly summarise here. First, the $\Sigma_\mathrm{SFR}$ map of JO201 was convolved with the \HI\ beam and regridded to the 5\arcsec\ pixels scale of the \HI\ map. This step was necessary to be able to make a direct comparison between the $\Sigma_\mathrm{SFR}$ and \HI\ map. Pixels in both maps were tagged as either belonging to the disc or tail (see Sec.\,\ref{j201HIresults} for details on how we defined the disc). We do note however that the tagging of JO201 pixels as disc or tail might be uncertain given the projection effects.

The disc and tail pixels were then compared to those of field galaxies extracted from The \HI\ Nearby Galaxy Survey (THINGS; \citealp{Bigiel2008}, \citealp{Bigiel2010}). We note that the $\Sigma_\mathrm{SFR}$ of THINGS galaxies are based on the combined FUV and 24$\mu$m emission but this is in good agreement with the H$\alpha$ emission (see Fig\,8 in \citealp{Bigiel2008}). Before we could compare our data with the control sample, we convolved the THINGS SFRD maps from their native spatial resolution of 750\,pc to the same physical resolution of JO201 -- 23 \,kpc. We then only selected galaxies from the THINGS sample which remained resolved after this convolution step which were NGC\,5055, NGC\,2841, NGC\,7331, NGC\,3198 and NGC\,3521.

\begin{figure}
  \centering
 \includegraphics[width=90mm, height=70mm]{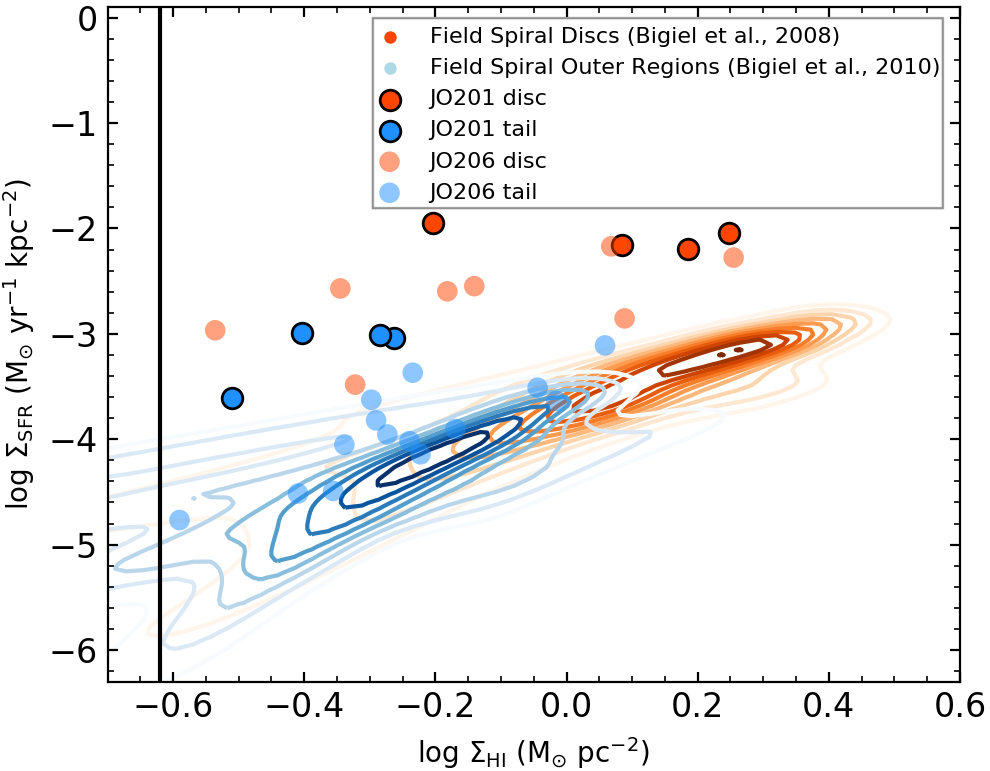}    
 \caption{The relation between the star formation rate density and H\textsc{i} surface density. The red and blue points are the main galaxy body and tail of JO201, respectively. Corresponding pixels of the JO206 are plotted in the same colour but more transparent. The pixels are plotted independently per beam to avoid showing correlated pixel values in a beam. Orange density contours are the inner regions (discs) of field spiral galaxies selected from the THINGS sample from \citealp{Bigiel2008} convolved with the H\textsc{i} beam. Light blue density contours represent the outer regions of spiral galaxies in the field \citep{Bigiel2010} also convolved with the H\textsc{i} beam. The solid vertical line indicates our general H\textsc{i} sensitivity limit.}\label{pixelSFR201}
\end{figure}

The $\Sigma_\mathrm{SFR}$ in relation to the \HI\ surface density for JO201 and JO206 is shown in Fig.\,\ref{pixelSFR201}. Pixels belonging to the discs and tails of both jellyfish galaxies and their THINGS counterparts (discs inner and outer regions) are shown in orange and blue, respectively. The $\Sigma_\mathrm{SFR}$ and \HI\ surface density values in the discs of JO201 and JO206 are deprojected for galaxy inclination. We note that the physical resolution of JO201 and that of JO206 are comparable, we therefore only show THINGS galaxies convolved with the \HI\ PSF JO201.

From this plot it is evident that both the disc and tail of these galaxies are forming stars at higher rate for a given \HI\ surface density compared to THINGS galaxies. For both, JO201 and JO206, the discs show star formation rate that is about 10 times higher on average than that of normal galaxies. However, as stated in \citet{Ramatsoku2019} these are regions where the \HI\ column density peaks at n$_{\rm H\textsc{i}} > 1.2 \times 10^{20}$ atoms cm$^{-2}$ ($\sim$0.9 M$_{\odot}$\,pc$^{-2}$). In the tails JO201 appears to have slightly higher star formation rate on average (3 times higher) at a given \HI\ surface density compared to the field counterparts. 

The physical reasons for the observed enhanced SFE in the disc and tail of JO201 and JO206 may be explained by combining our \HI\ data with Fig.\,7 in \citet{Moretti2018}, which shows their SFR as a function of H$_{2}$ from APEX data. From that plot the SFR of these galaxies appears to be low for the amount of H$_2$ detected. In our study we find that the SFR is high for the amount of \HI\ detected. This implies that the H$_{2}$/\HI\ ratio is high for these galaxies. A similar abundance of H$_2$ and deficiency of \HI\ (from upper limits) has also been reported in ESO137-001 which is another jellyfish galaxy that is not in the GASP sample (\citealp{Jachym2014}, \citeyear{Jachym2019}). The cause of this high ratio is currently not known and will be investigated in future papers. An interesting related observation is that, in a recent study of the magnetic field in one of the GASP jellyfish galaxies (e.g. Muller et al, submitted) finds magnetic field lines and synchrotron emission in the disc and tail of JO206. It is, therefore, possible that there is some shielding of the gas by the magnetic field which prevents the gas from evaporating into the hot ICM thus continue to form stars at a high rate even in the tails.

\subsection{Comparing the physical and environmental properties of JO201 and JO206}
In the previous sections we discussed the \HI\ and star formation properties of JO201 and JO206. Both galaxies show a strong displacement of the \HI\ relative to the stellar body (on the sky and/or in velocity); they have comparable stellar masses and \HI\ deficiencies, and they are forming stars at a similar rate, which is enhanced compared to that of galaxies with similar stellar and \HI\ mass. The interaction with the host cluster is thus having a similar effect on JO201 and JO206. This may seem strange given that the two galaxies are nearly identical, first in-fallers in two very different clusters (\citealp{Bellhouse2017}, \citealp{Poggianti2017}, \citealp{Jaffe2018}, \citealp{Bellhouse2019}; see also Table\,\ref{SummProp}). JO201 resides in a much more massive and denser cluster than JO206, and thus is expected to experience stronger ram pressure effects than the latter. In this section we attempt to understand whether the similarity between the two galaxies can at least in principle be explained despite the large difference between their host clusters.

\begin{table}
\caption {A comparison of physical and environmental properties of JO201 and JO206.}\label{SummProp}
\begin{tabular}{llllll}
\hline
Properties &  JO201   &  JO206   \\  
  
\hline\hline
M$_{\ast}$ (\MSUN)                       &  $3.6 \times10^{10}$ & $8.5 \times 10^{10}$ \\
M$_{\rm HI}$ (\MSUN)                   &  $1.7 \times 10^{9}$  & $3.2 \times 10^{9}$   \\                          
M$_{\rm HI}$/M$_{\ast}$                &  0.047                         &  0.037                      \\
\HI\ in the tail (\MSUN)                    & $0.4 \times 10^{9}$  & $1.8 \times 10^{9}$ \\
$\mu_{\ast}$ (\MSUN\ kpc$^{-2}$) &  $1.8 \times 10^{8}$  & $4.2 \times 10^{8}$   \\
SFR (\MSUN\,yr$^{-1}$)                 &   6.1                           &   5.6                           \\
$\tau_{\rm dep}$ (\HI) (Gyr)            &   0.27                           &   0.54                           \\
Host cluster                                     &  A85                             & IIZw108                    \\
M$_{\rm 200}$ (\MSUN)                 & $1.6 \times 10^{15}$   &  $2.0 \times 10^{14}$   \\
R$_{\rm 200}$ (Mpc)                     & 2.4                    & 1.4                     \\  
$\rho_{0}$  (kg\,m$^{-3}$)               & $2.6 \times 10^{-23}$   & $2.7 \times 10^{-24}$ \\
r$_{\rm c}$ (kpc)                  &  82                     & 261                             \\
$\beta_{\rm cl}$                               &  0.532                  & 0.662                          \\
$\sigma_{\rm cl}$ (\kms)                 &   982                                &   575                          \\
$v_{\rm gal}$ (\kms)                        &  3.4$\sigma_{\rm cl}$  & 1.14$\sigma_{\rm cl}$    \\
r$^{\rm proj}$ w.r.t cluster (kpc)       &  360                              &  351                          \\

\hline\hline\\
\end{tabular}
\end{table}

Throughout this study we have assumed instantaneous ram pressure following the \citet{Gunn1972} description of a homogeneous and symmetrical ICM. It is formulated as $P_\mathrm{ram} = \rho_{\rm \scriptscriptstyle ICM}v_\mathrm{gal}^{2}$, where $v_{\rm gal}$, is the differential velocity of the galaxy with respect to the cluster and $\rho_{\rm \scriptscriptstyle ICM}$ is the ICM density, which is parametrised with a $\beta$-model, $\rho_{\rm \scriptscriptstyle ICM} = \rho_0[1 + (r_{\rm gal}/r_{\rm c})^2]^{-3\beta/2}$ \citep{Cavaliere1976}. For this density profile $\rho_0$ is the central density, $r_c$ is the core radius of the cluster, $\beta$\ is the slope parameter and $r_\mathrm{gal}$ is the projected distance of the galaxy from the cluster centre. 

Using the aforementioned descriptions and the observed (i.e. projected) $r_\mathrm{gal}$ and $v_\mathrm{gal}$ of JO201 and JO206, we measure $P_{\rm ram}$ = $2.6 \times 10^{-11}$ Nm$^{-2}$ and $4.2 \times 10^{-13}$ Nm$^{-2}$, respectively. These estimates confirm our earlier statement that we expect stronger ram pressure effects on JO201 than JO206, at odds with the similarities between the two galaxies. 
We recall, however, that JO206 is falling into IIZw108 along the plane of the sky and consequently, its observed velocity is only a lower limit on its true 3D orbital velocity ($v^\mathrm{3D}$) around the cluster, while the projected distance is likely close to its true 3D orbital radius (i.e. $r^\mathrm{proj}_\mathrm{JO206}$ $\approx$ $r^\mathrm{3D}_\mathrm{JO206}$). JO201 on the other hand is falling into A\,85 along the line-of-sight such that $v^\mathrm{proj}_\mathrm{JO201}$ $\approx$ $v^\mathrm{3D}_\mathrm{JO201}$, while its projected distance from the cluster is a lower limit on its true 3D orbital radius.

In what follows we try to establish whether the large difference in the ram pressure estimates for JO201 and JO206 is thus a result of projection effects causing an underestimation of $r^\mathrm{3D}_\mathrm{JO201}$ and/or $v^\mathrm{3D}_\mathrm{JO206}$ -- while we assume that $r^\mathrm{3D}_\mathrm{JO206}$ and $v^\mathrm{3D}_\mathrm{JO201}$ are equal to their measured, projected values.

\begin{figure}
  \hspace*{-0.4cm}
 \includegraphics[width=96mm, height=73mm]{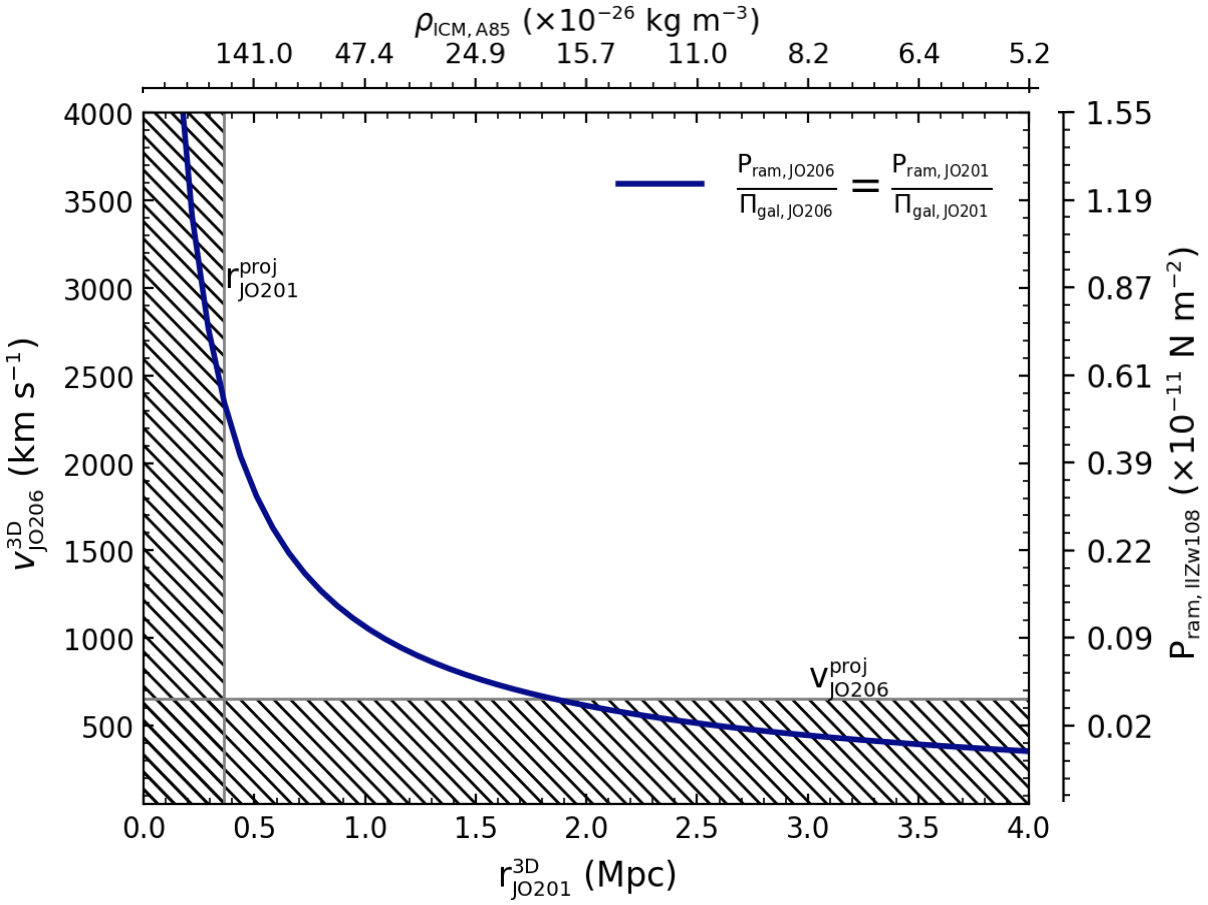}    
 \caption{Toy model showing what 3D velocity JO206 needs to have within the IIZw108 cluster in order to experience the same $P_{\rm ram}$ as JO201 (normalised by the gravitational restoring force $\Pi_\mathrm{gal}$), as a function of 3D distance of JO201 from the centre of the A 85 cluster. The blue curve indicates equal ram pressure strengths by the ICM of the clusters hosting JO201 (A\,85) and JO206 (IIZw108) relative to the galaxies' restoring force. Hatched vertical and horizontal regions represent forbidden regions given the measured, projected cluster-centric radius and velocity of JO201 and JO206, respectively. The quantities $r^\mathrm{3D}_\mathrm{JO206}$ and $v^\mathrm{3D}_\mathrm{JO201}$ are kept constant at their measured, projected values of 655.5 \kms\ and 360 kpc, respectively.}\label{RamPressureStrength}
\end{figure}

In Fig\,\ref{RamPressureStrength} we use a toy model to investigate what 3D velocities JO206 needs to have in IIZw108 so that it experiences the same P$_{\rm ram}$ (normalised by gravitational restoring force) as JO201 as a function of 3D distances of JO201 from the centre of its host cluster A\,85. We built this plot by varying $\rho_{\rm \scriptscriptstyle ICM}$ with $r^\mathrm{3D}_\mathrm{JO201}$ according to the $\beta$-model of A85.

We compare the ram pressure strengths with the restoring gravitational force per unit area of these galaxies defined as, $\prod_\mathrm{\rm gal} = 2\pi G\Sigma_\mathrm{g} \Sigma_\mathrm{s}$, where $G$ is the gravitational constant, $\Sigma_\mathrm{g}$ and $\Sigma_\mathrm{s}$ are the gas and stellar surface densities assuming an exponential profile of $\Sigma = \Sigma_{0}e^{-r_t/R_d}$ where $r_t$ is the radial distance from the galaxy centre, and $R_\mathrm{d}$ is the scale-length of the stellar disc. For this toy model we fix $\prod_{\rm gal}$ at a radial distance of $r_\mathrm{t} = 2R_\mathrm{d}$ which corresponds approximately to the distance from the galaxy centre at which the stripping conditions ($P_\mathrm{ram}/\prod_\mathrm{gal}$ > 1) are met (see \citealp{Poggianti2017}, \citealp{Bellhouse2019}).

Fig\,\ref{RamPressureStrength} shows the curve of equal ratio $P_\mathrm{ram}/\Pi_\mathrm{gal}$ for a range of 3D velocities and radii for JO206 and JO201, respectively. 
From this plot we deduce and conclude that there is a significantly large range of reasonable values of $v^\mathrm{3D}_\mathrm{JO206}$ and $r^\mathrm{3D}_\mathrm{JO201}$ of JO201 and JO206 to explain the similar \HI\ and star formation properties given the similarity of their stellar mass and morphology. We note that, assuming a \citealp{Navarro1996} potential for A\,85, the distance of JO201 from the cluster centre should be approximately $r^\mathrm{proj}_\mathrm{JO201}$ < $r^\mathrm{3D}_\mathrm{JO201}$ < 0.6 \ R$_{\rm 200} \sim 1.4$ Mpc for it to remain bound to the A\,85 cluster. Thus there is a region of Fig. 8 where JO201 and JO206 can experience similar ram-pressure stripping and JO201 is bound to its host cluster.

\section{Summary}\label{summary}
Within the context of the GASP survey we have examined the \HI\ gas phase of the jellyfish galaxy, JO201. The galaxy is a member of the massive cluster A\,85 with total mass of M$_{200} \sim 1.6 \times 10^{15}$\MSUN\ and  a velocity dispersion, $\sigma_{cl} \sim\ 982 \pm 55$\kms\ \citep{Moretti2017}. It is falling into the centre of this cluster along the line-of-sight with a velocity of $ = 3.4\sigma_{cl}$, and has long H$\alpha$ tails resulting from the ram pressure exerted by the cluster ICM. In this paper we studied its \HI\ content in relation to the star formation activity, and compared it to JO206 a similar jellyfish which resides in a significantly smaller cluster. Here we summarise our findings:

\begin{itemize}

\item{For JO201 we measure a total \HI\ mass of M$_{\rm HI} = 1.65 \times 10^{9}$ \MSUN. The galaxy is \HI\ deficient because this \HI\ mass is $\sim$60\% lower than expected based on its stellar mass and surface density. In \citet{Ramatsoku2019}, we reported a similar \HI\ deficiency of $\sim$50\% for JO206.}

\item{The \HI\ projected spatial distribution of JO201 is slightly perturbed but mostly close to the stellar disc. We find only a short \HI\ tail coinciding with the clumpy 40\,kpc long H$\alpha$ tail. No \HI\ is detected along the longer $\sim100$\,kpc long diffuse H$\alpha$, although we cannot rule out the presence of dense \HI\ clumps of similar size as the H$\alpha$ clumps, whose signal would be diluted below detectability due to the large \HI\ PSF. The close confinement of the \HI\ to the disc is probably due to projection effect since we also find that all of the \HI\ is strongly redshifted relative to the stars (100\,\kms\ or more) and is likely already outside of the stellar body in 3D.}\\

\item{An examination of the star formation rate in relation to the \HI\ content shows that JO201 has an enhanced SFR compared to field galaxies of the same stellar mass for its observed \HI\ content. We find that the time it would take to deplete its \HI\ through star formation at its current rate is $\sim$0.27\,Gyr (similarly that of JO206 - 0.5\,Gyr) which is much shorter than that of spiral galaxies in the field ($\sim$2\,Gyr; \citealp{Leroy2008}). Given its observed SFR, JO201 would need to have ten times its current \HI\ mass in order to be a `normal' spiral.}\\

\item{We conducted the resolved pixel-by-pixel analysis of the \HI\ surface density and the star formation rate density of JO201 for both the tail and disc. These properties were then compared with those of field galaxies from the THINGS. We found that similarly to JO206, the star formation efficiency of JO201 is about 10 times higher compared to that of field galaxies everywhere within the \HI\ distribution.}\\

\end{itemize}
~
Generally we observe similar physical properties for JO201 and JO206 based on their \HI\ mass, stellar masses and star formation rates. This would mean that the galaxies are experiencing similar ram pressure effects. This finding may appear inconsistent with the different environments in which these galaxies lie. 
However, we showed that when projection effects are taken into account the 3D position and velocity of the two galaxies within their respective (and very different) host clusters may well result in comparable ram pressure strengths relative to their restoring forces. The similarities between JO201 and JO206 in terms of \HI\ and SFR properties can therefore be explained despite the fact that they live in very different host clusters.

\begin{acknowledgements}
We thank the anonymous referee for the useful comments and suggestions. We also wish to thank Toby Brown for his invaluable help with the scaling relations. This project has received funding from the European Research Council (ERC) under the European Union's Horizon 2020 research and innovation programme grant agreement no. 679627 and no.833824, project name FORNAX and GASP, respectively. We acknowledge funding from the agreement ASI-INAF n.2017-14-H.0, as well as from the INAF main-stream funding programme. M.~R's research is supported by the SARAO HCD programme via the "New Scientific Frontiers with Precision Radio Interferometry" research group grant. M.~R. acknowledges support from the Italian Ministry of Foreign Affairs and International Cooperation (MAECI Grant Number ZA18GR02) and the South African Department of Science and Technology's National Research Foundation (DST-NRF Grant Number 113121) as part of the ISARP RADIOSKY2020 Joint Research Scheme. B.~V. and M.~G. also acknowledge the Italian PRIN-Miur 2017 (PI A. Cimatti). Y.~J. acknowledges financial support from CONICYT PAI (Concurso Nacional de Insercion en la Academia 2017), No. 79170132 and FONDECYT Iniciaci\'{o}n 2018 No. 11180558. M.~V. acknowledges support by the Netherlands Foundation for Scientific Research (NWO) through VICI grant 016.130.338. This work is based on observations collected at the European Organisation for Astronomical Research in the southern hemisphere under ESO programme 196.B-0578 (PI B.M. Poggianti). This work made use of THINGS, 'The \HI\ Nearby Galaxy Survey' \citep{Walter2008}. This paper makes use of the VLA data (Project code: VLA/17A-293). The National Radio Astronomy Observatory is a facility of the National Science Foundation operated under cooperative agreement by Associated Universities, Inc. (Part of) the data published here have been reduced using the CARAcal pipeline, partially supported by BMBF project 05A17PC2 for D-MeerKAT. Information about CARAcal can be obtained online under the \url{https://caracal.readthedocs.io/en/latest/}.

\end{acknowledgements}

\bibliographystyle{aa} 
\bibliography{J201GaspRefv2.bib} 

\begin{thebibliography}{78}
\expandafter\ifx\csname natexlab\endcsname\relax\def\natexlab#1{#1}\fi

\bibitem[{{Abramson} \& {Kenney}(2014)}]{Abramson2014}
{Abramson}, A. \& {Kenney}, J. D.~P. 2014, \aj, 147, 63

\bibitem[{{Abramson} {et~al.}(2011){Abramson}, {Kenney}, {Crowl}, {Chung}, {van
  Gorkom}, {Vollmer}, \& {Schiminovich}}]{Abramson2011}
{Abramson}, A., {Kenney}, J.~D.~P., {Crowl}, H.~H., {et~al.} 2011, \aj, 141,
  164

\bibitem[{{Bellhouse} {et~al.}(2017){Bellhouse}, {Jaff{\'e}}, {Hau}, {McGee},
  {Poggianti}, {Moretti}, {Gullieuszik}, {Bettoni}, {Fasano}, {D'Onofrio},
  {Fritz}, {Omizzolo}, {Sheen}, \& {Vulcani}}]{Bellhouse2017}
{Bellhouse}, C., {Jaff{\'e}}, Y.~L., {Hau}, G.~K.~T., {et~al.} 2017, \apj, 844,
  49

\bibitem[{{Bellhouse} {et~al.}(2019){Bellhouse}, {Jaff{\'e}}, {McGee},
  {Poggianti}, {Smith}, {Tonnesen}, {Fritz}, {Hau}, {Gullieuszik}, {Vulcani},
  {Fasano}, {Moretti}, {George}, {Bettoni}, {D'Onofrio}, {Omizzolo}, \&
  {Sheen}}]{Bellhouse2019}
{Bellhouse}, C., {Jaff{\'e}}, Y.~L., {McGee}, S.~L., {et~al.} 2019, \mnras,
  485, 1157

\bibitem[{{Bigiel} {et~al.}(2010){Bigiel}, {Leroy}, {Walter}, {Blitz},
  {Brinks}, {de Blok}, \& {Madore}}]{Bigiel2010}
{Bigiel}, F., {Leroy}, A., {Walter}, F., {et~al.} 2010, \aj, 140, 1194

\bibitem[{{Bigiel} {et~al.}(2008){Bigiel}, {Leroy}, {Walter}, {Brinks}, {de
  Blok}, {Madore}, \& {Thornley}}]{Bigiel2008}
{Bigiel}, F., {Leroy}, A., {Walter}, F., {et~al.} 2008, \aj, 136, 2846

\bibitem[{{Biviano} {et~al.}(2017){Biviano}, {Moretti}, {Paccagnella},
  {Poggianti}, {Bettoni}, {Gullieuszik}, {Vulcani}, {Fasano}, {D'Onofrio},
  {Fritz}, \& {Cava}}]{Biviano2017}
{Biviano}, A., {Moretti}, A., {Paccagnella}, A., {et~al.} 2017, \aap, 607, A81

\bibitem[{{Boissier} {et~al.}(2012){Boissier}, {Boselli}, {Duc}, {Cortese},
  {van Driel}, {Heinis}, {Voyer}, {Cucciati}, {Ferrarese}, {C{\^o}t{\'e}},
  {Cuillandre}, {Gwyn}, \& {Mei}}]{Boissier2012}
{Boissier}, S., {Boselli}, A., {Duc}, P.-A., {et~al.} 2012, \aap, 545, A142

\bibitem[{{Boselli} {et~al.}(2016){Boselli}, {Cuillandre}, {Fossati},
  {Boissier}, {Bomans}, {Consolandi}, {Anselmi}, {Cortese}, {C{\^o}t{\'e}},
  {Durrell}, {Ferrarese}, {Fumagalli}, {Gavazzi}, {Gwyn}, {Hensler}, {Sun}, \&
  {Toloba}}]{Boselli2016}
{Boselli}, A., {Cuillandre}, J.~C., {Fossati}, M., {et~al.} 2016, \aap, 587,
  A68

\bibitem[{{Boselli} \& {Gavazzi}(2006)}]{Boselli2006}
{Boselli}, A. \& {Gavazzi}, G. 2006, \pasp, 118, 517

\bibitem[{{Boselli} {et~al.}(1997){Boselli}, {Gavazzi}, {Lequeux}, {Buat},
  {Casoli}, {Dickey}, \& {Donas}}]{Boselli1997}
{Boselli}, A., {Gavazzi}, G., {Lequeux}, J., {et~al.} 1997, \aap, 327, 522

\bibitem[{{Bravo-Alfaro} {et~al.}(1997){Bravo-Alfaro}, {Szomoru}, {Cayatte},
  {Balkowski}, \& {Sancisi}}]{Bravo-Alfaro1997}
{Bravo-Alfaro}, H., {Szomoru}, A., {Cayatte}, V., {Balkowski}, C., \&
  {Sancisi}, R. 1997, \aaps, 126, 537

\bibitem[{{Brown} {et~al.}(2015){Brown}, {Catinella}, {Cortese}, {Kilborn},
  {Haynes}, \& {Giovanelli}}]{Brown2015}
{Brown}, T., {Catinella}, B., {Cortese}, L., {et~al.} 2015, \mnras, 452, 2479

\bibitem[{{Catinella} {et~al.}(2010){Catinella}, {Schiminovich}, {Kauffmann},
  {Fabello}, {Wang}, {Hummels}, {Lemonias}, {Moran}, {Wu}, {Giovanelli},
  {Haynes}, {Heckman}, {Basu-Zych}, {Blanton}, {Brinchmann}, {Budav{\'a}ri},
  {Gon{\c c}alves}, {Johnson}, {Kennicutt}, {Madore}, {Martin}, {Rich},
  {Tacconi}, {Thilker}, {Wild}, \& {Wyder}}]{Catinella2010}
{Catinella}, B., {Schiminovich}, D., {Kauffmann}, G., {et~al.} 2010, \mnras,
  403, 683

\bibitem[{{Cava} {et~al.}(2009){Cava}, {Bettoni}, {Poggianti}, {Couch},
  {Moles}, {Varela}, {Biviano}, {D'Onofrio}, {Dressler}, {Fasano}, {Fritz},
  {Kj{\ae}rgaard}, {Ramella}, \& {Valentinuzzi}}]{Cava2009}
{Cava}, A., {Bettoni}, D., {Poggianti}, B.~M., {et~al.} 2009, \aap, 495, 707

\bibitem[{{Cavaliere} \& {Fusco-Femiano}(1976)}]{Cavaliere1976}
{Cavaliere}, A. \& {Fusco-Femiano}, R. 1976, \aap, 500, 95

\bibitem[{{Chabrier}(2003)}]{Chabrier2003}
{Chabrier}, G. 2003, \pasp, 115, 763

\bibitem[{{Chung} {et~al.}(2009){Chung}, {van Gorkom}, {Kenney}, {Crowl}, \&
  {Vollmer}}]{Chung2009}
{Chung}, A., {van Gorkom}, J.~H., {Kenney}, J.~D.~P., {Crowl}, H., \&
  {Vollmer}, B. 2009, \aj, 138, 1741

\bibitem[{{Cortese} {et~al.}(2006){Cortese}, {Gavazzi}, {Boselli}, {Franzetti},
  {Kennicutt}, {O'Neil}, \& {Sakai}}]{Cortese2006}
{Cortese}, L., {Gavazzi}, G., {Boselli}, A., {et~al.} 2006, \aap, 453, 847

\bibitem[{{Cortese} {et~al.}(2007){Cortese}, {Marcillac}, {Richard},
  {Bravo-Alfaro}, {Kneib}, {Rieke}, {Covone}, {Egami}, {Rigby}, {Czoske}, \&
  {Davies}}]{Cortese2007}
{Cortese}, L., {Marcillac}, D., {Richard}, J., {et~al.} 2007, \mnras, 376, 157

\bibitem[{{Cowie} \& {Songaila}(1977)}]{Cowie1977}
{Cowie}, L.~L. \& {Songaila}, A. 1977, \nat, 266, 501

\bibitem[{{Cramer} {et~al.}(2019){Cramer}, {Kenney}, {Sun}, {Crowl}, {Yagi},
  {J{\'a}chym}, {Roediger}, \& {Waldron}}]{Cramer2019}
{Cramer}, W.~J., {Kenney}, J.~D.~P., {Sun}, M., {et~al.} 2019, \apj, 870, 63

\bibitem[{{Deb} {et~al.}(2020){Deb}, {Verheijen}, {Gullieuszik}, {Poggianti},
  {van Gorkom}, {Ramatsoku}, {Serra}, {Moretti}, {Vulcani}, {Bettoni}, {Jaffe},
  {Tonnesen}, \& {Fritz}}]{Deb2020}
{Deb}, T., {Verheijen}, M. A.~W., {Gullieuszik}, M., {et~al.} 2020, arXiv
  e-prints, arXiv:2004.04754

\bibitem[{{Di Teodoro} \& {Fraternali}(2015)}]{DiTeodoro2015}
{Di Teodoro}, E.~M. \& {Fraternali}, F. 2015, \mnras, 451, 3021

\bibitem[{{Doyle} \& {Drinkwater}(2006)}]{Doyle2006}
{Doyle}, M.~T. \& {Drinkwater}, M.~J. 2006, \mnras, 372, 977

\bibitem[{{Ebeling} {et~al.}(2014){Ebeling}, {Stephenson}, \&
  {Edge}}]{Ebeling2014}
{Ebeling}, H., {Stephenson}, L.~N., \& {Edge}, A.~C. 2014, \apjl, 781, L40

\bibitem[{{Fasano} {et~al.}(2006){Fasano}, {Marmo}, {Varela}, {DOnofrio},
  {Poggianti}, {Moles}, {Pignatelli}, {Bettoni}, {Kj{\ae}rgaard}, {Rizzi},
  {Couch}, \& {Dressler}}]{Fasano2006}
{Fasano}, G., {Marmo}, C., {Varela}, J., {et~al.} 2006, \aap, 445, 805

\bibitem[{{Fasano} {et~al.}(2012){Fasano}, {Vanzella}, {Dressler}, {Poggianti},
  {Moles}, {Bettoni}, {Valentinuzzi}, {Moretti}, {D'Onofrio}, {Varela},
  {Couch}, {Kj{\ae}rgaard}, {Fritz}, {Omizzolo}, \& {Cava}}]{Fasano2012}
{Fasano}, G., {Vanzella}, E., {Dressler}, A., {et~al.} 2012, \mnras, 420, 926

\bibitem[{{Fossati} {et~al.}(2016){Fossati}, {Fumagalli}, {Boselli}, {Gavazzi},
  {Sun}, \& {Wilman}}]{Fossati2016}
{Fossati}, M., {Fumagalli}, M., {Boselli}, A., {et~al.} 2016, \mnras, 455, 2028

\bibitem[{{Fumagalli} {et~al.}(2014){Fumagalli}, {Fossati}, {Hau}, {Gavazzi},
  {Bower}, {Sun}, \& {Boselli}}]{Fumagalli2014}
{Fumagalli}, M., {Fossati}, M., {Hau}, G.~K.~T., {et~al.} 2014, \mnras, 445,
  4335

\bibitem[{{Gavazzi} {et~al.}(2018){Gavazzi}, {Consolandi}, {Gutierrez},
  {Boselli}, \& {Yoshida}}]{Gavazzi2018}
{Gavazzi}, G., {Consolandi}, G., {Gutierrez}, M.~L., {Boselli}, A., \&
  {Yoshida}, M. 2018, \aap, 618, A130

\bibitem[{{Gavazzi} {et~al.}(2008){Gavazzi}, {Giovanelli}, {Haynes}, {Fabello},
  {Fumagalli}, {Kent}, {Koopmann}, {Brosch}, {Hoffman}, {Salzer}, \&
  {Boselli}}]{Gavazzi2008}
{Gavazzi}, G., {Giovanelli}, R., {Haynes}, M.~P., {et~al.} 2008, \aap, 482, 43

\bibitem[{{George} {et~al.}(2018){George}, {Poggianti}, {Gullieuszik},
  {Fasano}, {Bellhouse}, {Postma}, {Moretti}, {Jaff{\'e}}, {Vulcani},
  {Bettoni}, {Fritz}, {C{\^o}t{\'e}}, {Ghosh}, {Hutchings}, {Mohan},
  {Sreekumar}, {Stalin}, {Subramaniam}, \& {Tandon}}]{George2018}
{George}, K., {Poggianti}, B.~M., {Gullieuszik}, M., {et~al.} 2018, \mnras,
  479, 4126

\bibitem[{{Gullieuszik} {et~al.}(2015){Gullieuszik}, {Poggianti}, {Fasano},
  {Zaggia}, {Paccagnella}, {Moretti}, {Bettoni}, {D'Onofrio}, {Couch},
  {Vulcani}, {Fritz}, {Omizzolo}, {Baruffolo}, {Schipani}, {Capaccioli}, \&
  {Varela}}]{Gullieuszik2015}
{Gullieuszik}, M., {Poggianti}, B., {Fasano}, G., {et~al.} 2015, \aap, 581, A41

\bibitem[{{Gunn} \& {Gott}(1972)}]{Gunn1972}
{Gunn}, J.~E. \& {Gott}, III, J.~R. 1972, \apj, 176, 1

\bibitem[{{Haynes} {et~al.}(1984){Haynes}, {Giovanelli}, \&
  {Chincarini}}]{Haynes1984}
{Haynes}, M.~P., {Giovanelli}, R., \& {Chincarini}, G.~L. 1984, \araa, 22, 445

\bibitem[{{Hester} {et~al.}(2010){Hester}, {Seibert}, {Neill}, {Wyder}, {Gil de
  Paz}, {Madore}, {Martin}, {Schiminovich}, \& {Rich}}]{Hester2010}
{Hester}, J.~A., {Seibert}, M., {Neill}, J.~D., {et~al.} 2010, \apjl, 716, L14

\bibitem[{{Huang} {et~al.}(2012){Huang}, {Haynes}, {Giovanelli}, \&
  {Brinchmann}}]{Huang2012}
{Huang}, S., {Haynes}, M.~P., {Giovanelli}, R., \& {Brinchmann}, J. 2012, \apj,
  756, 113

\bibitem[{{Jachym} {et~al.}(2014){Jachym}, {Combes}, {Cortese}, {Sun}, \&
  {Kenney}}]{Jachym2014}
{Jachym}, P., {Combes}, F., {Cortese}, L., {Sun}, M., \& {Kenney}, J.~D.~P.
  2014, \apj, 792, 11

\bibitem[{{J{\'a}chym} {et~al.}(2019){J{\'a}chym}, {Kenney}, {Sun}, {Combes},
  {Cortese}, {Scott}, {Sivanandam}, {Brinks}, {Roediger}, {Palou{\v{s}}}, \&
  {Fumagalli}}]{Jachym2019}
{J{\'a}chym}, P., {Kenney}, J. D.~P., {Sun}, M., {et~al.} 2019, \apj, 883, 145

\bibitem[{{Jaffe} {et~al.}(2018){Jaffe}, {Poggianti}, {Moretti}, {Gullieuszik},
  {Smith}, {Vulcani}, {Fasano}, {Fritz}, {Tonnesen}, {Bettoni}, {Hau},
  {Biviano}, {Bellhouse}, \& {McGee}}]{Jaffe2018}
{Jaffe}, Y.~L., {Poggianti}, B.~M., {Moretti}, A., {et~al.} 2018, \mnras, 476,
  4753

\bibitem[{{Jaff{\'e}} {et~al.}(2015){Jaff{\'e}}, {Smith}, {Candlish},
  {Poggianti}, {Sheen}, \& {Verheijen}}]{Jaffe2015}
{Jaff{\'e}}, Y.~L., {Smith}, R., {Candlish}, G.~N., {et~al.} 2015, \mnras, 448,
  1715

\bibitem[{{Kapferer} {et~al.}(2009){Kapferer}, {Sluka}, {Schindler}, {Ferrari},
  \& {Ziegler}}]{Kapferer2009}
{Kapferer}, W., {Sluka}, C., {Schindler}, S., {Ferrari}, C., \& {Ziegler}, B.
  2009, \aap, 499, 87

\bibitem[{{Kenney} {et~al.}(2008){Kenney}, {Tal}, {Crowl}, {Feldmeier}, \&
  {Jacoby}}]{Kenney2008}
{Kenney}, J.~D.~P., {Tal}, T., {Crowl}, H.~H., {Feldmeier}, J., \& {Jacoby},
  G.~H. 2008, \apjl, 687, L69

\bibitem[{{Leroy} {et~al.}(2008){Leroy}, {Walter}, {Brinks}, {Bigiel}, {de
  Blok}, {Madore}, \& {Thornley}}]{Leroy2008}
{Leroy}, A.~K., {Walter}, F., {Brinks}, E., {et~al.} 2008, \aj, 136, 2782

\bibitem[{{Martinsson} {et~al.}(2016){Martinsson}, {Verheijen}, {Bershady},
  {Westfall}, {Andersen}, \& {Swaters}}]{Martinsson2016}
{Martinsson}, T.~P.~K., {Verheijen}, M.~A.~W., {Bershady}, M.~A., {et~al.}
  2016, \aap, 585, A99

\bibitem[{{McPartland} {et~al.}(2016){McPartland}, {Ebeling}, {Roediger}, \&
  {Blumenthal}}]{McPartland2016}
{McPartland}, C., {Ebeling}, H., {Roediger}, E., \& {Blumenthal}, K. 2016,
  \mnras, 455, 2994

\bibitem[{{Moretti} {et~al.}(2017){Moretti}, {Gullieuszik}, {Poggianti},
  {Paccagnella}, {Couch}, {Vulcani}, {Bettoni}, {Fritz}, {Cava}, {Fasano},
  {D'Onofrio}, \& {Omizzolo}}]{Moretti2017}
{Moretti}, A., {Gullieuszik}, M., {Poggianti}, B., {et~al.} 2017, \aap, 599,
  A81

\bibitem[{{Moretti} {et~al.}(2018){Moretti}, {Paladino}, {Poggianti},
  {D'Onofrio}, {Bettoni}, {Gullieuszik}, {Jaff{\'e}}, {Vulcani}, {Fasano},
  {Fritz}, \& {Torstensson}}]{Moretti2018}
{Moretti}, A., {Paladino}, R., {Poggianti}, B.~M., {et~al.} 2018, \mnras, 480,
  2508

\bibitem[{{Navarro} {et~al.}(1996){Navarro}, {Frenk}, \& {White}}]{Navarro1996}
{Navarro}, J.~F., {Frenk}, C.~S., \& {White}, S. D.~M. 1996, \apj, 462, 563

\bibitem[{{Nulsen}(1982)}]{Nulsen1982}
{Nulsen}, P.~E.~J. 1982, \mnras, 198, 1007

\bibitem[{{Owen} {et~al.}(2006){Owen}, {Keel}, {Wang}, {Ledlow}, \&
  {Morrison}}]{Owen2006}
{Owen}, F.~N., {Keel}, W.~C., {Wang}, Q.~D., {Ledlow}, M.~J., \& {Morrison},
  G.~E. 2006, \aj, 131, 1974

\bibitem[{{Owers} {et~al.}(2012){Owers}, {Couch}, {Nulsen}, \& {Rand
  all}}]{Owers2012}
{Owers}, M.~S., {Couch}, W.~J., {Nulsen}, P. E.~J., \& {Rand all}, S.~W. 2012,
  \apjl, 750, L23

\bibitem[{{Perley} {et~al.}(2009){Perley}, {Napier}, {Jackson}, {Butler},
  {Carlson}, {Fort}, {Dewdney}, {Clark}, {Hayward}, {Durand}, {Revnell}, \&
  {McKinnon}}]{Perley2009}
{Perley}, R., {Napier}, P., {Jackson}, J., {et~al.} 2009, IEEE Proceedings, 97,
  1448

\bibitem[{{Poggianti} {et~al.}(2016){Poggianti}, {Fasano}, {Omizzolo},
  {Gullieuszik}, {Bettoni}, {Moretti}, {Paccagnella}, {Jaff{\'e}}, {Vulcani},
  {Fritz}, {Couch}, \& {D'Onofrio}}]{Poggianti2016}
{Poggianti}, B.~M., {Fasano}, G., {Omizzolo}, A., {et~al.} 2016, \aj, 151, 78

\bibitem[{{Poggianti} {et~al.}(2019){Poggianti}, {Gullieuszik}, {Tonnesen},
  {Moretti}, {Vulcani}, {Radovich}, {Jaff{\'e}}, {Fritz}, {Bettoni},
  {Franchetto}, {Fasano}, {Bellhouse}, \& {Omizzolo}}]{Poggianti2019}
{Poggianti}, B.~M., {Gullieuszik}, M., {Tonnesen}, S., {et~al.} 2019, \mnras,
  482, 4466

\bibitem[{{Poggianti} {et~al.}(2017{\natexlab{a}}){Poggianti}, {Jaff{\'e}},
  {Moretti}, {Gullieuszik}, {Radovich}, {Tonnesen}, {Fritz}, {Bettoni},
  {Vulcani}, {Fasano}, {Bellhouse}, {Hau}, \& {Omizzolo}}]{Poggianti2017nat}
{Poggianti}, B.~M., {Jaff{\'e}}, Y.~L., {Moretti}, A., {et~al.}
  2017{\natexlab{a}}, \nat, 548, 304

\bibitem[{{Poggianti} {et~al.}(2017{\natexlab{b}}){Poggianti}, {Moretti},
  {Gullieuszik}, {Fritz}, {Jaff{\'e}}, {Bettoni}, {Fasano}, {Bellhouse}, {Hau},
  {Vulcani}, {Biviano}, {Omizzolo}, {Paccagnella}, {D'Onofrio}, {Cava},
  {Sheen}, {Couch}, \& {Owers}}]{Poggianti2017}
{Poggianti}, B.~M., {Moretti}, A., {Gullieuszik}, M., {et~al.}
  2017{\natexlab{b}}, \apj, 844, 48

\bibitem[{{Quilis} {et~al.}(2000){Quilis}, {Moore}, \& {Bower}}]{Quilis2000}
{Quilis}, V., {Moore}, B., \& {Bower}, R. 2000, Science, 288, 1617

\bibitem[{{Radovich} {et~al.}(2019){Radovich}, {Poggianti}, {Jaff{\'e}},
  {Moretti}, {Bettoni}, {Gullieuszik}, {Vulcani}, \& {Fritz}}]{Radovich2019}
{Radovich}, M., {Poggianti}, B., {Jaff{\'e}}, Y.~L., {et~al.} 2019, \mnras,
  486, 486

\bibitem[{{Ramatsoku} {et~al.}(2019){Ramatsoku}, {Serra}, {Poggianti},
  {Moretti}, {Gullieuszik}, {Bettoni}, {Deb}, {Fritz}, {van Gorkom},
  {Jaff{\'e}}, {Tonnesen}, {Verheijen}, {Vulcani}, {Hugo}, {J{\'o}zsa},
  {Maccagni}, {Makhathini}, {Ramaila}, {Smirnov}, \& {Thorat}}]{Ramatsoku2019}
{Ramatsoku}, M., {Serra}, P., {Poggianti}, B.~M., {et~al.} 2019, \mnras, 487,
  4580

\bibitem[{{Saintonge} {et~al.}(2016){Saintonge}, {Catinella}, {Cortese},
  {Genzel}, {Giovanelli}, {Haynes}, {Janowiecki}, {Kramer}, {Lutz},
  {Schiminovich}, {Tacconi}, {Wuyts}, \& {Accurso}}]{Saintonge2016}
{Saintonge}, A., {Catinella}, B., {Cortese}, L., {et~al.} 2016, \mnras, 462,
  1749

\bibitem[{{Scott} {et~al.}(2010){Scott}, {Bravo-Alfaro}, {Brinks}, {Caretta},
  {Cortese}, {Boselli}, {Hardcastle}, {Croston}, \& {Plauchu}}]{Scott2010}
{Scott}, T.~C., {Bravo-Alfaro}, H., {Brinks}, E., {et~al.} 2010, \mnras, 403,
  1175

\bibitem[{{Serra} {et~al.}(2015){Serra}, {Westmeier}, {Giese}, {Jurek},
  {Fl{\"o}er}, {Popping}, {Winkel}, {van der Hulst}, {Meyer}, {Koribalski},
  {Staveley-Smith}, \& {Courtois}}]{Serra2015}
{Serra}, P., {Westmeier}, T., {Giese}, N., {et~al.} 2015, \mnras, 448, 1922

\bibitem[{{Smith} {et~al.}(2010){Smith}, {Lucey}, {Hammer}, {Hornschemeier},
  {Carter}, {Hudson}, {Marzke}, {Mouhcine}, {Eftekharzadeh}, {James},
  {Khosroshahi}, {Kourkchi}, \& {Karick}}]{RSmith2010}
{Smith}, R.~J., {Lucey}, J.~R., {Hammer}, D., {et~al.} 2010, \mnras, 408, 1417

\bibitem[{{Steinhauser} {et~al.}(2016){Steinhauser}, {Schindler}, \&
  {Springel}}]{Steinhauser2016}
{Steinhauser}, D., {Schindler}, S., \& {Springel}, V. 2016, \aap, 591, A51

\bibitem[{{Sun} {et~al.}(2007){Sun}, {Donahue}, \& {Voit}}]{Sun2007}
{Sun}, M., {Donahue}, M., \& {Voit}, G.~M. 2007, \apj, 671, 190

\bibitem[{{Varela} {et~al.}(2009){Varela}, {D'Onofrio}, {Marmo}, {Fasano},
  {Bettoni}, {Cava}, {Couch}, {Dressler}, {Kj{\ae}rgaard}, {Moles},
  {Pignatelli}, {Poggianti}, \& {Valentinuzzi}}]{Varela2009}
{Varela}, J., {D'Onofrio}, M., {Marmo}, C., {et~al.} 2009, \aap, 497, 667

\bibitem[{{Venkatapathy} {et~al.}(2017){Venkatapathy}, {Bravo-Alfaro}, {Mayya},
  {Lobo}, {Durret}, {Gamez}, {Valerdi}, {Granados-Contreras}, \&
  {Navarro-Poupard}}]{Venkatapathy2017}
{Venkatapathy}, Y., {Bravo-Alfaro}, H., {Mayya}, Y.~D., {et~al.} 2017, \aj,
  154, 227

\bibitem[{{Verdugo} {et~al.}(2015){Verdugo}, {Combes}, {Dasyra}, {Salom{\'e}},
  \& {Braine}}]{Verdugo2015}
{Verdugo}, C., {Combes}, F., {Dasyra}, K., {Salom{\'e}}, P., \& {Braine}, J.
  2015, \aap, 582, A6

\bibitem[{{Vollmer} {et~al.}(2001){Vollmer}, {Cayatte}, {van Driel}, {Henning},
  {Kraan-Korteweg}, {Balkowski}, {Woudt}, \& {Duschl}}]{Vollmer2001}
{Vollmer}, B., {Cayatte}, V., {van Driel}, W., {et~al.} 2001, \aap, 369, 432

\bibitem[{{Vollmer} {et~al.}(2012){Vollmer}, {Wong}, {Braine}, {Chung}, \&
  {Kenney}}]{Vollmer2012}
{Vollmer}, B., {Wong}, O.~I., {Braine}, J., {Chung}, A., \& {Kenney}, J.~D.~P.
  2012, \aap, 543, A33

\bibitem[{{Vulcani} {et~al.}(2018){Vulcani}, {Poggianti}, {Gullieuszik},
  {Moretti}, {Tonnesen}, {Jaff{\'e}}, {Fritz}, {Fasano}, \&
  {Bettoni}}]{Vulcani2018}
{Vulcani}, B., {Poggianti}, B.~M., {Gullieuszik}, M., {et~al.} 2018, \apjl,
  866, L25

\bibitem[{{Walter} {et~al.}(2008){Walter}, {Brinks}, {de Blok}, {Bigiel},
  {Kennicutt}, {Thornley}, \& {Leroy}}]{Walter2008}
{Walter}, F., {Brinks}, E., {de Blok}, W.~J.~G., {et~al.} 2008, \aj, 136, 2563

\bibitem[{{Wang} {et~al.}(2016){Wang}, {Koribalski}, {Serra}, {van der Hulst},
  {Roychowdhury}, {Kamphuis}, \& {Chengalur}}]{Wang2016}
{Wang}, J., {Koribalski}, B.~S., {Serra}, P., {et~al.} 2016, \mnras, 460, 2143

\bibitem[{{Yagi} {et~al.}(2010){Yagi}, {Yoshida}, {Komiyama}, {Kashikawa},
  {Furusawa}, {Okamura}, {Graham}, {Miller}, {Carter}, {Mobasher}, \&
  {Jogee}}]{Yagi2010}
{Yagi}, M., {Yoshida}, M., {Komiyama}, Y., {et~al.} 2010, \aj, 140, 1814

\bibitem[{{Yoon} {et~al.}(2017){Yoon}, {Chung}, {Smith}, \&
  {Jaff{\'e}}}]{Yoon2017}
{Yoon}, H., {Chung}, A., {Smith}, R., \& {Jaff{\'e}}, Y.~L. 2017, \apj, 838, 81

\bibitem[{{Yoshida} {et~al.}(2008){Yoshida}, {Yagi}, {Komiyama}, {Furusawa},
  {Kashikawa}, {Koyama}, {Yamanoi}, {Hattori}, \& {Okamura}}]{Yoshida2008}
{Yoshida}, M., {Yagi}, M., {Komiyama}, Y., {et~al.} 2008, \apj, 688, 918

\end{thebibliography}

\end{document}